\documentclass{article}

\usepackage{amsthm}
\usepackage{amsmath}
\usepackage{amssymb}
\usepackage{graphicx}
\usepackage{hyperref}
\usepackage{eprint}
\usepackage{dcolumn}
\usepackage{bm}
\usepackage[osf]{mathpazo} 
\usepackage{url}
\urldef{\mailsa}\path|{alfred.hofmann, ursula.barth, ingrid.beyer, christine.guenther,|
\urldef{\mailsb}\path|frank.holzwarth, anna.kramer, erika.siebert-cole, lncs}@springer.com|

\newcommand{\targfix}{\qw {\xy {<0em,0em> \ar @{ - } +<.39em,0em>
\ar @{ - } -<.39em,0em> \ar @{ - } +
<0em,.39em> \ar @{ - }
-<0em,.39em>},<0em,0em>*{\rule{.01em}{.01em}}*+<.8em>\frm{o}
\endxy}}

\input{Qcircuit}

\newcommand{\normtwo}{\frac{1}{\sqrt{2}}}

\newcommand{\email}[1]{\href{mailto:#1}{#1}}
\newtheorem{definition}{Definition}
\theoremstyle{plain} \newtheorem{lemma}{Lemma}

\begin{document}

\title{A 2D Nearest-Neighbor Quantum Architecture for Factoring in Polylogarithmic Depth}


%
%
\author{Paul Pham\\
University of Washington\\
Quantum Theory Group\\
Box 352350, Seattle, WA 98195, USA,\\
\email{ppham@cs.washington.edu},\\
\url{http://www.cs.washington.edu/homes/ppham/}
\and
Krysta M. Svore\\
Microsoft Research\\
Quantum Architectures and Computation Group\\
One Microsoft Way, Redmond, WA 98052, USA\\
\email{ksvore@microsoft.com},\\
\url{http://research.microsoft.com/en-us/people/ksvore/}
}

\maketitle

\begin{abstract}
We contribute a 2D nearest-neighbor
quantum architecture for Shor's algorithm to factor an $n$-bit number in $O(\log^2(n))$ depth.
Our implementation uses
parallel phase estimation,
constant-depth fanout and teleportation,
and
constant-depth carry-save modular addition.
We derive upper bounds on the circuit resources of our architecture under a
new 2D nearest-neighbor model
which allows a classical controller and parallel, communicating
modules. We also contribute a novel constant-depth circuit for
unbounded quantum \emph{unfanout} in our new model.
Finally, we provide a comparison to all previous nearest-neighbor factoring
implementations.  
Our circuit results in an exponential improvement in nearest-neighbor circuit depth at the cost of a polynomial increase in circuit size and width.
\end{abstract}

\section{Introduction}
\label{sec:intro}

Shor's factoring algorithm is a central result in quantum computing, with an
exponential speed-up over the best known classical algorithm \cite{Shor1994}.
As the most notable example of a quantum-classical complexity separation, much
effort has been devoted to implementations of factoring on a
realistic architectural model of a quantum computer
\cite{Beauregard2002,Kutin2006,VanMeter2006,VanMeter2005,VanMeterIL2005}.
We can bridge the gap between
the theoretical algorithm and a physical implementation by describing
the layout and interactions of qubits at an intermediate,
architectural level of abstraction.
This gives us a model for measuring circuit resources and their tradeoffs.
In this work, we contribute a circuit implementation for prime
factorization of an $n$-bit integer
on a two-dimensional architecture that allows concurrent (parallel) two-qubit operations
between neighboring qubits, an omnipresent classical controller, and
modules which are allowed to teleport qubits to each other. We call this new
model \textsc{2D CCNTCM}.
We show that our circuit construction is asymptotically more efficient in circuit depth than previous state-of-the-art techniques for nearest-neighbor
architectures, achieving a depth of $O(\log^2 n)$, a size of
$O(n^4)$, and a width of $O(n^4)$ qubits, as detailed in Table
\ref{tab:results} of Section \ref{sec:results}.

Our technique hinges on several key building blocks.
Section \ref{sec:bg} introduces quantum architectural models, circuit
resources, and constant-depth communication techniques due to
\cite{Harrow2012,Rosenbaum2012}, including a circuit for
constant-depth unfanout which is unique to the current work.
Section \ref{sec:related} places our work in the context of existing
results.
In Section \ref{sec:csa}, we provide a self-contained pedagogical review
of the carry-save technique and encoding.
In Section \ref{sec:csa-mod-add} we modify and extend the carry-save technique to a 2D
modular adder,
which we then use as a basis for a modular multiplier
(Section \ref{sec:csa-mod-mult}) and a modular exponentiator
(Section \ref{sec:modexp}).
For each building block, we provide numerical upper bounds for the
required circuit resources.
Finally, we compare our asymptotic circuit resource usage
with other factoring implementations.

\section{Background}
\label{sec:bg}

Quantum architecture is the design of physical qubit layouts
and their allowed interactions to execute
quantum algorithms efficiently in time, space, and other
resources.
In this paper, we focus on designing a realistic nearest-neighbor circuit for running
Shor's factoring algorithm on two-dimensional
architectural models of a physical quantum device with nearest-neighbor
interactions.

\subsection{Architectural Models and Circuit Resources}
\label{subsec:models}

Following Van Meter and Itoh \cite{VanMeter2005},
we distinguish between a model and an architectural implementation as follows.
A \emph{model} is a set of constraints and rules for the placement and
interaction of qubits.
An \emph{architecture} (or interchangeably, an \emph{implementation} 
or a \emph{circuit}) is a particular
spatial layout of qubits (as a graph of vertices) and allowed interactions (edges between the vertices),
following the constraints of a given model. In this section, we describe
several models which try to incorporate resources of physical interest from
experimental work. We also introduce a new model,
\textsc{2D CCNTCM}, which we will use to analyze our current circuit.

The most general model is called Abstract Concurrent (\textsc{AC})
and allows arbitrary, long-range interactions between any qubits and concurrent
operation of quantum gates.
This corresponds to a complete graph with an edge between every pair of nodes.
It is the model assumed in most quantum algorithms.

A more specialized model restricts interactions to nearest-neighbor, two-qubit,
concurrent gates (\textsc{NTC}) in a regular one-dimensional chain (1D NTC),
which is sometimes called linear nearest-neighbor (\textsc{LNN}).
This corresponds to a line graph. This is a more realistic model than
\textsc{AC}, but correspondingly, circuits in this model may incur greater
resource overheads.

To relieve movement congestion,
we can consider a two-dimensional regular grid
(2D NTC), where each
qubit has four planar neighbors, and 
there is an extra degree of freedom over the 1D model
in which to move data.
In this paper, we extend the \textsc{2D NTC} model in three ways.
The first two extensions are described in Section \ref{subsec:2dccntc},
and the third extension is described in Section \ref{subsec:2dccntcm}.

\subsection{\textsc{2D CCNTC}: Two-Dimensional Nearest-Neighbor Two-Qubit Concurrent Gates with Classical Controller}
\label{subsec:2dccntc}

The first extension allows arbitrary planar graphs
with bounded degree, rather than a regular square lattice.
Namely, we assume qubits lie in a plane and edges are not allowed to intersect.
All qubits are accessible from above
or below by control and measurement apparatus.
Whereas 2D NTC conventionally assumes each qubit
has four neighbors, we consider up to six neighbors in a roughly hexagonal
layout. The edge length in this model is no more than twice the edge length
in a regular 2D NTC lattice. The second extension is the realistic assumption
that classical control (CC) can
access every qubit in parallel, and we do not count these classical
resources in our implementation since they are polynomially bounded. The
classical controllers
correspond to fast digital computers which are
available in actual experiments and are necessary for constant-depth
communication in the next section.

We call an AC or NTC model augmented by these two extensions
\textsc{CCAC} and \textsc{CCNTC}, respectively. Before we describe the
third extension, let us formalize our model for \textsc{2D CCNTC}, with definitions that are (asymptotically) equivalent to those in 
\cite{Rosenbaum2012}.

\begin{definition}
A 2D CCNTC architecture consists of

\begin{itemize}
\item a quantum computer $QC$ which is represented by a planar graph $(V,E)$. A
node $v \in V$ represents a qubit which is acted upon in a circuit, and an
undirected edge $(u,v) \in E$ represents 
an allowed two-qubit interaction between qubits $u,v \in V$. Each node has
degree at most $6$.
\item a universal gate set $\mathcal{G} = \{X, Z, H, T, T^{\dagger}, CNOT, MeasureZ\}$.

\item a deterministic machine (classical controller) $CC$ that applies a sequence
of concurrent gates in each of $D$ timesteps.
\item In timestep $i$, $CC$ applies a set of
gates $G_i = \{g_{i,j} \in \mathcal{G} \}$.
Each $g_{i,j}$ operates in one of the following two ways:
\begin{enumerate}
\item It is a single-qubit gate from $\mathcal{G}$ acting on a single qubit $v_{i,j} \in V$
\item
It is the gate CNOT from $\mathcal{G}$ acting on two qubits $v^{(1)}_{i,j}, v^{(2)}_{i,j} \in V$ where
$(v^{(1)}_{i,j}, v^{(2)}_{i,j}) \in E$
\end{enumerate}
All the $g_{i,j}$ can only operate on
disjoint qubits for a given timestep $i$. We define the support of $G_i$
as $V_i$, the set of all qubits acted upon during timestep $i$.

\begin{equation}
V_i = \bigcup_{j: g_{i,j} \in G_i} v_{i,j} \cup v^{(1)}_{i,j} \cup v^{(2)}_{i,j}
\end{equation}

\end{itemize}
\end{definition}

We can then define the three conventional circuit resources in this model.

\begin{description}
\item[circuit depth ($D$):] the number of concurrent timesteps.
\item[circuit size ($S$):] the total number of non-identity gates applied
from $\mathcal{G}$, equal to $\sum_{i=1}^D |G_i|$.
\item[circuit width ($W$):] the total number of qubits operated upon by
any gate, including inputs, outputs, and ancillae. It is equal to $| \bigcup_{i=1}^D V_i|$.
\end{description}

We observe that the following relationship holds between the circuit resources.
The circuit size is bounded above by
the product of circuit depth and circuit width, since in the worst case,
every qubit is acted upon by a gate for every timestep of a circuit.
The circuit depth is also bounded above by the size, since in the worst case,
every gate is executed serially without any concurrency.

\begin{equation}
D \le S \le D\cdot W
\label{eqn:depth-width}
\end{equation}

\begin{figure*}[tb!]
\begin{center}
\begin{displaymath}
\begin{array}{ccc}
\Qcircuit @C=1em @R=1em {
& \qw & \multimeasureD{1}{\mbox{Bell}} & \cw & \rstick{j} \\
& \qw & \ghost{\mbox{Bell}}            & \cw & \rstick{k}
}
& \qquad \equiv \qquad &
\Qcircuit @C=1em @R=1em {
& \qw & \ctrl{1} & \qw & \gate{H} & \qw & \meter & \cw & \rstick{j} \\
& \qw & \targfix & \qw & \qw      & \qw & \meter & \cw & \rstick{k}
}
\end{array}
\end{displaymath}
\centerline{}
\caption{A circuit for measurement in the Bell state basis.}
\label{fig:bell-measure}
\end{center}\end{figure*}
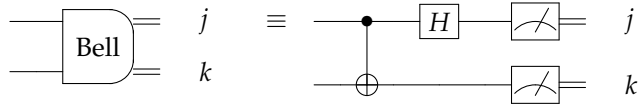

The set $\mathcal{G}$ includes measurement in the $Z$ basis, which is
actually not a unitary operation but which may be slower than unitary
operations in actual practice \cite{DiVincenzo2007}.
Therefore we count it in our resource
estimates.
All other gates
in $\mathcal{G}$ form a universal set of unitary
gates \cite{Kitaev2002}.
 In this paper we
will treat the operations in $\mathcal{G}$ as \emph{elementary gates}.
We can also define a Bell basis measurement using operations
from $\mathcal{G}$. A circuit performing this measurement is shown
in Figure \ref{fig:bell-measure} and has depth $4$,
size $4$, and width $2$.

The third extension to our model, and the most significant, is to consider
multiple disconnected planar graphs, each of which is a 2D CCNTC
architecture. This is described in more detail in the next section.

\subsection{\textsc{2D CCNTCM}: Two-Dimensional Nearest-Neighbor Two-Qubit Concurrent Gates with Classical Controller and Modules}
\label{subsec:2dccntcm}

A single, contiguous
2D lattice which contains an entire quantum architecture which may be prohibitively large to manufacture. In practice,
scalable experiments will probably use many
smaller quantum computers which communicate by means of shared
entanglement \cite{Monroe2012}.
We call these individual machines \emph{modules}, each of
which is a self-contained \textsc{2D CCNTC} lattice. This should not be
confused with the word ``modular'' as in ``modular arithmetic'' or as
referring to the modulus $m$ which we are trying to factor.

We treat these modules
and teleportations between them as nodes and edges, respectively,
in a higher-level planar graph. The teleportations each transmit one qubit
from one module to another, from any location within the source module
to any location within the destination module, making use of the
omnipresent classical controller. The modules can be arbitrarily far
apart physically, but they have bounded-degree connectivity with other
modules, and their edges are planar (they cannot intersect).

\begin{figure}[btp!]
\begin{center}
\includegraphics[width=4in]{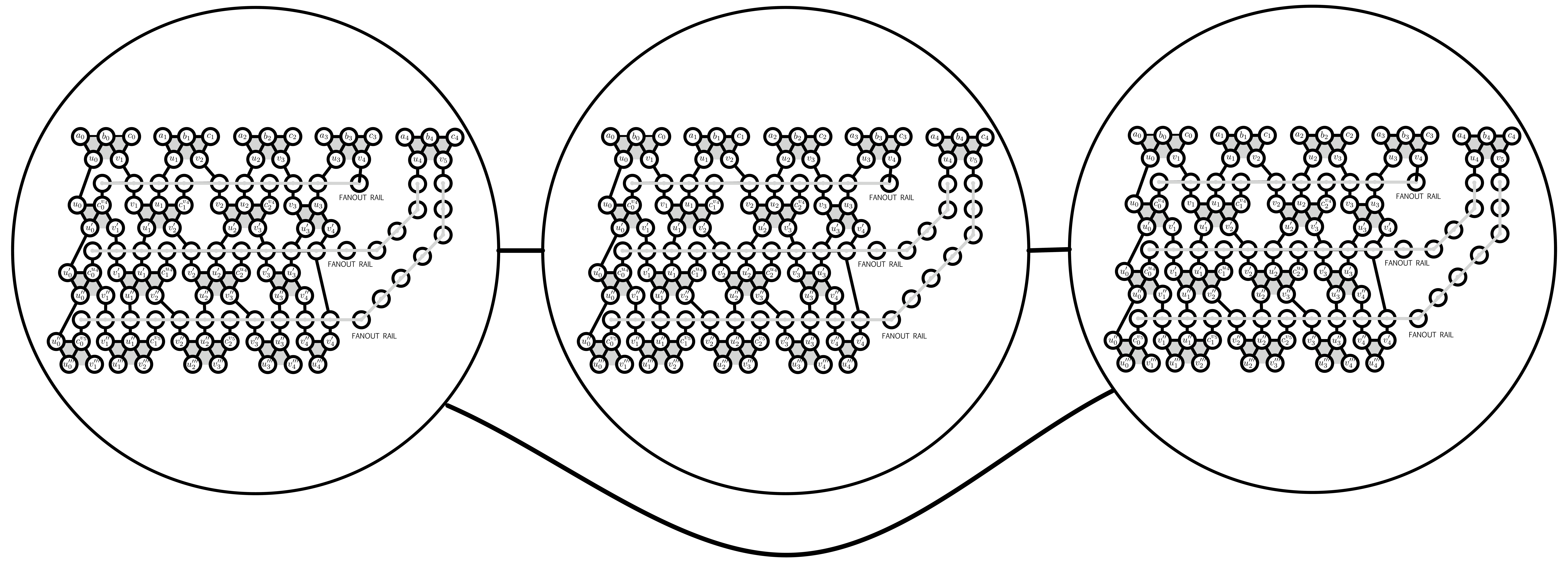}
\end{center}
\caption{Three modules in the \textsc{2D CCNTCM} model}
\label{fig:modules}
\end{figure}

A single module can be part of multiple teleportation operations in a single timestep, as long as they involve disjoint qubits within the module.
We justify this assumption in that it is
possible to establish entanglement between multiple
quantum computers
in parallel. We call this new model \textsc{2D CCNTCM},
and we argue that is captures the essential aspects of 2D architectures
without being overly sensitive to the exact geometry of the lattices involved.
An graphic depiction of three modules in \textsc{2D CCNTCM} is shown in
Figure \ref{fig:modules}. Each module contains within it a
\textsc{2D CCNTC} lattice. We can equivalently consider the omnipresent,
single
classical controller as a collection of multiple classical controllers, one
for each module or teleportation operation, which can inter-communicate
classically and share a clock.

\begin{definition}
A \textsc{2D CCNTCM} architecture consists of

\begin{itemize}
\item a quantum computer $\overline{QC}$ which is represented by a planar graph $(\overline{V},\overline{E})$. A
node $\overline{v} \in \overline{V}$ represents a module, or a graph $(V,E)$
from a \textsc{2D CCNTC} architecture defined previously. It can have
unbounded degree.
An
undirected edge $(\overline{u},\overline{v}) \in \overline{E}$ represents an
allowed teleportation from any qubit in module $\overline{u}$ to
another qubit in module $\overline{v}$.
\item All modules are restricted to be linear in the number of their qubits:
$|V| = \Theta(n)$ for all $(V,E) \in \overline{V}$.
\item a universal gate set $\mathcal{G} = \{X, Z, H, T, T^{\dagger}, CNOT,
MeasureZ\}$
for the qubits \emph{within the same} modules which is the same as for \textsc{2D CCNTC},
and an additional operation $Teleport$ which only operates on qubits
\emph{in
different} modules.
\item a deterministic machine (classical controller) $\overline{CC}$ that applies a sequence
of concurrent gates in each of $D+\overline{D}$ timesteps.
This can be a separate classical controller
for every pair of modules.
\item In timestep $i$, $\overline{CC}$ applies
gates $G_i = \{g_{i,j} : g_{i,j} \in \mathcal{G} \lor g_{i,j} = Teleport \}$.
That is, there are two kinds of timesteps with respect to the kinds of gates
which operate within them.
\begin{enumerate}
\item In the first kind, gates are exclusively from $\mathcal{G}$, and
they operate within modules as described
for \textsc{2D CCNTC} above. We say there are $D$ such timesteps.
\item In the second kind, gates are exclusively $Teleport$ gates between two qubits $v^{(1)}_{i,j} \in \overline{v}_1$ and
$v^{(2)}_{i,j} \in \overline{v}_2$ for
(possibly non-distinct) modules $\overline{v}_1, \overline{v}_2 \in \overline{V}$.
Again, all such qubits much be distinct within a timestep.
We say there are $\overline{D}$ such timesteps.
\end{enumerate}

Again, we define the support of $G_i$
as $V_i$, the set of all qubits acted upon by any $g_{i.j}$, which
includes all the modules.
\begin{equation}
V_i = \bigcup_{j: g_{i,j} \in G_i} v_{i,j} \cup v^{(1)}_{i,j} \cup v^{(2)}_{i,j} 
\end{equation}

\end{itemize}
\end{definition}

We measure the efficiency of a circuit in this new module using not just
the three conventional circuit resources, but with three novel resources
based on modules.

\begin{description}

\item[module depth ($\overline{D}$):] the depth of consecutive teleportations between modules.
\item[module size ($\overline{S}$):] the number of total qubits teleported between any two modules over all timesteps.
\item[module width ($\overline{W}$):] the number of modules whose qubits are
acted upon during any timestep.

\end{description}



We note the following relationship between circuit width and
module width.

\begin{equation}
W = O(n\overline{W})
\label{eqn:module-width}
\end{equation}

This restriction imposes some locality on our model by constraining it to
nearest-neighbor gates within a linear-sized group of qubits, but allowing
it long-range teleportation to circumvent onerous geometric constraints.
Using the constant-depth communication in Section \ref{subsec:fanout}, and for
the specific case of factoring, we
can simulate arbitrary connectivity between modules with only a polynomial
increase in the module size and a constant increase in module depth.

\subsection{Circuit Resource Comparisons}

Counting gates from $\mathcal{G}$ as having unit size and unit depth
is
an overestimate compared to the model in \cite{Kutin2006}, in which a
two-qubit gate has unit size and unit depth and
absorbs the depth and size of any adjacent single-qubit gates. We intend
for this more pessimistic estimate to reflect the practical difficulties
in compiling these gates using a non-Clifford gate in a fault-tolerant way,
such as the $T$ gate or the Toffoli gate
\cite{Fowler2011}.

In both our resource counting method and that of \cite{Fowler2004,Kutin2006}, multiple gates acting on disjoint qubits
can occur in parallel during the same timestep. For each building block,
from modular addition to modular multiplication and finally to modular
exponentiation, we provide closed form equations upper-bounding the required circuit
resources as a function of $n$, the size of the modulus $m$ to be factored.
We will use the
term \emph{numerical upper bound} to distinguish these formulae from asymptotic
upper bounds.

It is possible to reduce the numerical constants with more detailed analysis,
which would be important for any physical implementation.
However, we have chosen instead to simplify the number of terms in the formulae
for the current work. We do not intend for these upper bounds to represent
the optimal or final work in this area.

The modular adder in Section \ref{sec:csa-mod-add} and its carry-save
subcomponents only occur within a single module, so we only give their
circuit resources in terms of circuit depth, circuit size, and circuit width. 
For the modular multiplier in
Section \ref{sec:csa-mod-mult} and the modular exponentiator in
Section \ref{sec:modexp}, we also give circuit resources in
terms of module depth, module size, and module width.

\subsection{Constant-depth Teleportation, Fanout, and Unfanout}
\label{subsec:fanout}

Communication, namely the \emph{moving} and \emph{copying} of quantum information, in nearest-neighbor quantum architectures is challenging.
In this section we quote known results for teleportation and
fanout in constant depth while also contributing a novel construction
for unfanout.

\begin{figure*}[tb!]
\begin{center}
\begin{displaymath}
\Qcircuit @C=1em @R=1em {
\lstick{\ket{\psi}}	& \qw      & \qw      & \qw & \qw & \qw & \qw & \qw                                          & \qw & \qw & \multimeasureD{1}{\mbox{Bell}} & \cw & \rstick{j_1} \\
\lstick{\ket{0}}    & \gate{H} & \ctrl{1} & \qw & \qw & \qw & \qw & \qw                                          & \qw & \qw & \ghost{\mbox{Bell}}            & \cw & \rstick{k_1} \\
\lstick{\ket{0}}    & \qw      & \targfix & \qw & \qw & \qw & \qw & \qw_{Z^{j_1}X^{k_1}\ket{\psi}}               & \qw & \qw & \multimeasureD{1}{\mbox{Bell}} & \cw & \rstick{j_2} \\
\lstick{\ket{0}}    & \gate{H} & \ctrl{1} & \qw & \qw & \qw & \qw & \qw                                          & \qw & \qw & \ghost{\mbox{Bell}}            & \cw & \rstick{k_2} \\
\lstick{\ket{0}}    & \qw      & \targfix & \qw & \qw & \qw & \qw & \qw_{Z^{j_2}Z^{j_1}X^{k_2}X^{k_1}\ket{\psi}} & \qw & \qw & \multimeasureD{1}{\mbox{Bell}} & \cw & \rstick{j_3} \\
\lstick{\ket{0}}    & \gate{H} & \ctrl{1} & \qw & \qw & \qw & \qw & \qw                                          & \qw & \qw & \ghost{\mbox{Bell}}            & \cw & \rstick{k_3} \\
\lstick{\ket{0}}    & \qw      & \targfix & \qw & \qw & \qw & \qw & \qw & \qw_{Z^{j_1}Z^{j_2}Z^{j_3}X^{k_3}X^{k_2}X^{k_1}\ket{\psi}} & \qw & \qw              & \qw & \qw \\
}
\end{displaymath}
\centerline{}
\caption{Constant-depth circuit based on \protect{\cite{Broadbent2007,Browne2009}} for teleportation over $n=5$ qubits \protect{\cite{Rosenbaum2012}}.}
\label{fig:cdt}
\end{center}\end{figure*}
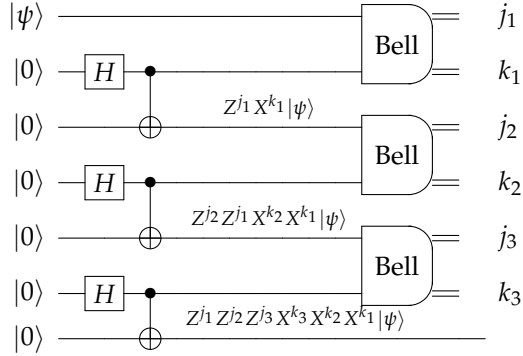

The first challenge of moving quantum information from one site to another over
arbitrarily long distances can be addressed by using
the constant-depth teleportation circuit
shown in Figure \ref{fig:cdt} due to Rosenbaum \cite{Rosenbaum2012}, illustrated using standard quantum circuit
notation \cite{Nielsen2000}. This requires the circuit resources shown in
Table \ref{tab:cd-resources}. The depth includes a layer of $H$ gates; a layer of CNOTs; an interleaved layer of Bell basis measurements; and two layers of
Pauli corrections ($X$ and $Z$ for each qubit), occurring concurrently with
resetting the $\ket{j}$ and $\ket{k}$ qubits back to $\ket{0}$.
These correction layers are not shown in the circuit.

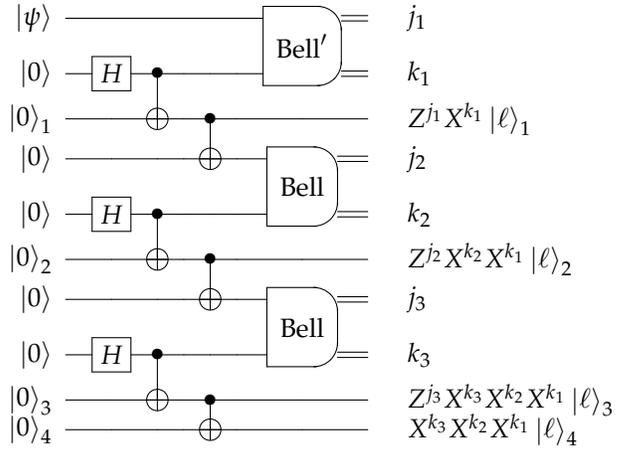
\begin{figure*}[tb!]
\begin{center}
\begin{displaymath}
\Qcircuit @C=1em @R=1em {
\lstick{\ket{\psi}}	& \qw      & \qw      & \qw & \qw & \qw & \multimeasureD{1}{\mbox{Bell}'} & \cw & \rstick{j_1} \\
\lstick{\ket{0}}    & \gate{H} & \ctrl{1} & \qw & \qw      & \qw & \ghost{\mbox{Bell}'}            & \cw & \rstick{k_1} \\
\lstick{\ket{0}_1}    & \qw      & \targfix & \qw & \ctrl{1} & \qw & \qw      & \qw & \rstick{Z^{j_1}X^{k_1}\ket{\ell}_1}\\
\lstick{\ket{0}}	& \qw      & \qw      & \qw & \targfix & \qw & \multimeasureD{1}{\mbox{Bell}} & \cw & \rstick{j_2} \\
\lstick{\ket{0}}    & \gate{H} & \ctrl{1} & \qw & \qw      & \qw & \ghost{\mbox{Bell}}           & \cw & \rstick{k_2} \\
\lstick{\ket{0}_2}    & \qw      & \targfix & \qw & \ctrl{1} & \qw & \qw      & \qw & \rstick{Z^{j_2}X^{k_2}X^{k_1}\ket{\ell}_2}\\
\lstick{\ket{0}}	& \qw      & \qw      & \qw & \targfix & \qw & \multimeasureD{1}{\mbox{Bell}} & \cw & \rstick{j_3} \\
\lstick{\ket{0}}    & \gate{H} & \ctrl{1} & \qw & \qw      & \qw & \ghost{\mbox{Bell}}           & \cw & \rstick{k_3} \\
\lstick{\ket{0}_3}    & \qw      & \targfix & \qw & \ctrl{1} & \qw & \qw      & \qw & \rstick{Z^{j_3}X^{k_3}X^{k_2} X^{k_1}\ket{\ell}_3}\\
\lstick{\ket{0}_4}	& \qw      & \qw      & \qw & \targfix & \qw & \qw      & \qw & \rstick{X^{k_3}X^{k_2} X^{k_1}\ket{\ell}_4}\\
}
\end{displaymath}
\centerline{}
\caption{Constant-depth circuits based on \protect{\cite{Broadbent2007,Browne2009}} for fanout \protect{\cite{Harrow2012}} of one qubit to $n=4$ entangled copies.}
\label{fig:cdf}
\end{center}\end{figure*}

Although general cloning is
impossible \cite{Nielsen2000}, the second challenge of copying information can be addressed by performing an unbounded quantum
fanout operation:
$\ket{x,y_1,\ldots,y_n} \rightarrow \ket{x,y_1\oplus x, \ldots, y_n\oplus x}$.
This is used in our arithmetic circuits when
a single qubit needs to control (be entangled with) a large quantum register
(called a \emph{fanout rail}).
We employ a constant-depth circuit due to insight from
measurement-based quantum computing \cite{Raussendorf2003}
that relies on the creation of an
$n$-qubit cat state \cite{Browne2009} which was communicated to
us by Harrow and Fowler \cite{Harrow2012}.

This circuit requires $O(1)$-depth, $O(n)$-size, and $O(n)$-width. Approximately
two-thirds of the ancillae are reusable and can be reset to $\ket{0}$ after
being measured. Numerical upper bounds are given in Table \ref{tab:cd-resources}.
The constant-depth fanout circuit is shown in Figure \ref{fig:cdf} for the case of fanning out a given single-qubit state
$\ket{\psi} = \alpha\ket{0} + \beta\ket{1}$ to four qubits.
The technique works by creating multiple small
cat states of a fixed size (in this case, three qubits), linking them
together into a larger cat state of unbounded size with Bell basis measurements,
and finally entangling them with the source qubit to be fanned out.
The qubits marked $\ket{\ell}$ are
entangled into the larger fanned out state given in Equation \ref{eqn:cat4}.
The Pauli corrections from the cat state creation are denoted by
$X^{k_2}$, $X^{k_3}$, $Z^{j_2}$ and $Z^{j_3}$ on qubits ending in
states $\ket{\ell}_1$, $\ket{\ell}_2$,
$\ket{\ell}_3$, and $\ket{\ell}_4$. The Pauli corrections
$X^{k_1}$ and $Z^{j_1}$ are from the Bell basis measurement
entangling the cat state with the source qubit (denoted $\text{Bell}'$).
\begin{equation}
Z_1^{j_1}X_1^{k_1}Z_2^{j_2}X_2^{k_2}X_2^{k_1}Z_{3}^{j_3}X_{3}^{k_3}X_{3}^{k_2}X_{3}^{k_1}X_{4}^{k_3}X_{4}^{k_2}X_{4}^{k_1}
\left(\alpha \ket{0}_1\ket{0}_2\ket{0}_3\ket{0}_4 + \beta \ket{1}_1\ket{1}_2\ket{1}_3\ket{1}_4 \right)
\label{eqn:cat4}
\end{equation}
The operators $X^k_i$ and $Z^j_{h}$ denote Pauli $X$ and $Z$ operators
on qubits $i$ and $h$, controlled by classical bits $k$ and $j$,
respectively. These corrections are enacted by the classical controller based on
the Bell measurement outcomes (not depicted).
Note the cascading nature of these corrections.
There can be up to
$n-1$ of these $X$ and $Z$
corrections on the same qubit, which can be simplified by the classical
controller to a single $X$ and $Z$ operation and then applied with a circuit of
depth 2 and size 2. Also, given the symmetric nature of the cat state, there
is an alternate set of Pauli corrections which would give the same state and
is of equal size to the corrections given above.

\begin{figure*}[tb!]
\begin{center}
\begin{displaymath}
\Qcircuit @C=1em @R=1em {
& \lstick{\ket{\ell}}	& \qw & \gate{H} & \qw & \ctrl{1} & \qw & \qw      & \qw &  \measureD{Z} & \cw & \rstick{j_1} & \\
& \lstick{\ket{\ell}}	& \qw & \gate{H} & \qw & \targfix & \qw & \ctrl{1} & \qw & \measureD{Z} & \cw & \rstick{j_2} & \\
& \lstick{\ket{\ell}}	& \qw & \gate{H} & \qw & \ctrl{1} & \qw & \targfix & \qw & \measureD{Z} & \cw & \rstick{j_3} & \\
& \lstick{\ket{\ell}}	& \qw & \gate{H} & \qw & \targfix & \qw & \ctrl{1} & \qw & \measureD{Z} & \cw & \rstick{j_4} & \\
& \lstick{\ket{\ell}}	& \qw & \gate{H} & \qw & \ctrl{1} & \qw & \targfix & \qw & \measureD{Z} & \cw & \rstick{j_5} & \\
& \lstick{\ket{\ell}}	& \qw & \gate{H} & \qw & \targfix & \qw & \ctrl{1} & \qw & \measureD{Z} & \cw & \rstick{j_6} & \\
& \lstick{\ket{\ell}}	& \qw & \gate{H} & \qw & \qw      & \qw & \targfix & \qw & \gate{H} & \qw & \rstick{Z^{j_2 \oplus j_4}(\alpha\ket{0} + \beta\ket{1})}
}
\end{displaymath}
\centerline{}
\caption{A novel, constant-depth circuit for unbounded quantum unfanout on
CCNTC, from the $7$-qubit entangled state $\alpha\ket{0}^{\otimes 7} + \beta\ket{1}^{\otimes 7}$ to the
target product state $(\alpha\ket{0} + \beta\ket{1})\otimes\ket{0}^{\otimes 6}$.}
\label{fig:cdu}
\end{center}\end{figure*}
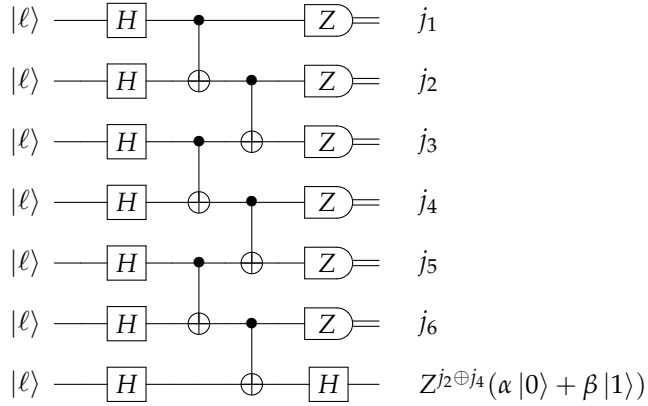

Reversing the fanout is an operation called \emph{unfanout}. Unfanout
takes as input 
the following entangled $n$-qubit state which is the result of a fanout.

\begin{equation}
\normtwo (\ket{0}^{\otimes n} + \ket{1}^{\otimes n})
\label{eqn:fanned-out}
\end{equation}

The output of unfanout, after Pauli corrections, is the product state
consisting of all $\ket{0}$'s except for a single target qubit $\alpha\ket{0} + \beta\ket{1}$, which is in the
same state as the original source qubit of the fanout.

In the model of \cite{Hoyer2002}, the fanout and unfanout were identical, elementary
operations. In CCNTC, the operations are not identical due to
the one-way nature of the measurement and the
constraints of NTC. In Figure \ref{fig:cdu}, we contribute a novel quantum circuit for unbounded quantum
unfanout for $n=7$ in constant depth on 2D CCNTC.
Note that the state in Equation \ref{eqn:fanned-out}
is completely symmetric in that all qubits are
equivalent entangled copies of each other. Therefore, the asymmetry 
of the final target qubit is entirely determined by the unfanout circuit,
which in this case selects the bottom qubit in the figure.

The initial fanned out state lives in a $2$-dimensional subspace. The
round of Hadamard gates increases its dimension to $2^n$, and the two
interleaved layers of CNOTs in a sense ``disentangle'' the qubits from
one another, up to a Pauli $Z$ correction. This correction, on the
final target qubit, is controlled by the parity of the classical measurements
on every ``even'' qubit ($j_2$ and $j_4$ in the figure), excluding the 
next-to-last qubit ($j_6$ in the figure). Each measurement projects the state of the
target qubit
into a subspace with half the dimension, so $n-1$ measurements project
the target qubit into a final $2$-dimensional subspace, which is the
qubit $\alpha\ket{0} + \beta\ket{1}$.

Although the circuit show works for odd $n$, we can easily take into
account even $n$ with an initial CNOT to ``uncopy'' one qubit from its
neighbors. The unfanout circuit in Figure \ref{fig:cdu} is the
functional inverse of the
fanout circuit in 
Figure \ref{fig:cdf}, but it relies on the fanned-out qubits
being teleported back into adjacent positions,
which is only possible in a 2D layout.
The target qubit of unfanout is usually chosen to be in the same location
as the source qubit of the corresponding fanout. 
The resources for unfanout are given in
Table \ref{tab:cd-resources}.

\begin{table}
\begin{displaymath}
\begin{tabular}{|c|c|c|c|}
\hline
\text{Circuit Name} & \text{Depth} & \text{Size} & \text{Width}\\
\hline
\text{Teleportation from Figure \ref{fig:cdt}} & 7 & 3n + 4 & n+1\\
\hline
\text{Fanout from Figure \ref{fig:cdf}} & 9 & 10n - 9 & 3n-1 \\
\hline
\text{Unfanout} & $ 6 $ & $ 3n+2 $ & $ n$ \\
\hline
\end{tabular}
\end{displaymath}
\centerline{}
\caption{Circuit resources for teleportation, fanout, and unfanout
(consisting of
alternating rounds of constant-depth teleportation and CNOT).}
\label{tab:cd-resources}
\end{table}

From an experimental perspective, it is physically efficient to create
a cat state in trapped ions using the M{\o}lmer-S{\o}rensen gate
\cite{Sorensen2000}\cite{Benhelm2008}. However, the fanout circuit for
the 2D CCNTCM model would still be useful for other technologies, such
as superconducting qubits on a 2D lattice.


\section{Related Work}
\label{sec:related}

Our work builds upon ideas in classical digital and reversible logic and their extension to quantum logic.
Any circuit implementation for Shor's algorithm requires a quantum adder.
Gossett proposed a quantum algorithm for addition using classical carry-save techniques to add
in constant-depth and multiply in logarithmic-depth, with a quadratic
cost in qubits (circuit width) \cite{Gossett1998}. The techniques relies on encoded addition, sometimes
called a 3-2 adder, and derives from classical Wallace trees \cite{Wallace1964}.

Takahashi and Kunihiro discovered a linear-depth
and linear-size quantum adder using zero ancillae \cite{Takahashi2005}.
They also developed an adder with tradeoffs between $O(n/d(n))$ ancillae and
$O(d(n))$-depth for $d(n) = \Omega(\log n)$ \cite{Takahashi2009}. 
Their approach assumes unbounded fanout, which had not previously been mapped to a
nearest-neighbor circuit until our present work.

Studies of architectural constraints, namely restriction to a 2D planar layout, 
were experimentally motivated. For example, these layouts correspond
to early ion trap proposals \cite{Kielpinski2002}
and were later analyzed at the level of physical qubits and error correction in the context of Shor's algorithm \cite{Kubi09}.
Choi and Van Meter designed one of the first adders targeted to a 2D architecture 
and showed it runs in $\Theta(\sqrt{n})$-depth on \textsc{2D NTC} \cite{Choi2010}
using $O(n)$-qubits with dedicated, special-purpose areas of a physical
circuit layout.

Modular exponentiation is a key component of quantum period-finding (QPF),
and its efficiency relies on that of its underlying adder implementation.
Since Shor's algorithm is a probabilistic algorithm, multiple rounds of
QPF are required to amplify success probability arbitrarily close to 1.
It suffices to determine the resources
required for a single round of QPF with a fixed, modest success probability
(in the current work, $3/4$).

The most common approach to QPF performs controlled
modular exponentiation followed by an inverse quantum Fourier transform
(QFT) \cite{Nielsen2000}. We will call this \emph{serial QPF}, which is
used by the following implementations.

Beauregard \cite{Beauregard2002}
constructs a cubic-depth quantum period-finder using only $2n+3$ qubits on
\textsc{AC}.
It combines the ideas of Draper's transform adder \cite{Draper2000},
Vedral et al.'s modular arithmetic blocks \cite{Vedral1996}, and a
semi-classical QFT.
This approach was subsequently adapted to \textsc{1D NTC} by Fowler, Devitt,
and Hollenberg
\cite{Fowler2004} to achieve resource counts for an $O(n^3)$-depth
quantum period-finder. Kutin \cite{Kutin2006} later improved this using
an idea from Zalka for approximate multipliers to produce a QPF circuit on
\textsc{1D NTC}
in $O(n^2)$-depth. Thus, there is only a constant overhead from
Zalka's own factoring implementation on \textsc{AC}, which also has
quadratic depth \cite{Zalka1998}.
Takahashi and Kunihiro extended their earlier $O(n)$-depth adder to a factoring
circuit in $O(n^3)$-depth with linear width \cite{Takahashi2006}.
Van Meter and Itoh explore many different approaches for serial QPF,
with their lowest achievable depth being $O(n^2\log n)$ with
$O(n^2)$ on \textsc{NTC} \cite{VanMeter2005}. Cleve and Watrous
calculate a factoring circuit depth of $O(\log^3 n)$ and corresponding
circuit size of $O(n^3)$ on \textsc{AC},
assuming an adder which has depth $O(\log n)$ and
$O(n)$ size and width. We beat this depth and provide a concrete
architectural implementation using an adder with $O(1)$-depth and $O(n)$
size and width.

In the current work, we assume that errors do not affect the storage of qubits
during the circuit's operation. An alternate approach is taken by
Miquel \cite{Miquel1996} and Garcia-Mata \cite{GarciaMata2007}, who both
numerically simulate Shor's algorithm for factoring specific
numbers to determine its sensitivity to errors. Beckman et al. provide a
concrete factoring implementation in ion traps with $O(n^3)$ depth and size and
$O(n)$ width \cite{Beckman1996}.

In all the previous works,
it is assumed that qubits are expensive (width) and that
execution time (depth) is not the limiting constraint.
We make the alternative assumption that ancillae are cheap and that fast classical control
is available which allows access to all qubits in parallel.
Therefore, we optimize circuit depth at the expense of width.
We compare our work primarily to Kutin's method \cite{Kutin2006}.

These works also rely on serial QPF which in turn relies on an inverse QFT.
On an AC architecture, even when approximating the (inverse) QFT by truncating two-qubit
$\pi/2^k$ rotations beyond $k = O(\log n)$, 
the depth is $O(n \log n)$ to factor an $n$-bit number.
To be implemented fault-tolerantly on a quantum device, rotations in the QFT must then be compiled into a discrete gate basis.
This requires at least a $O(\log(1/\epsilon))$ overhead in depth to approximate a rotation with precision $\epsilon$ \cite{Harrow02, Kitaev2002}.
We would like to avoid the use of a QFT due to its compilation overhead.

There is an alternative, parallel version of phase estimation 
\cite{Cleve2000,Kitaev2002}, which we call \emph{parallel QPF} (we refer the reader to Section 13 of \cite{Kitaev2002} for details), which decreases depth in exchange
for increased width and additional classical post-processing.
This eliminates the need to do an inverse QFT.
We develop a nearest-neighbor factoring circuit based on parallel QPF and our proposed 2D quantum arithmetic circuits.
We show that it is asymptotically more efficient than the serial QPF method. 
We compare the circuit resources required by our work with existing serial QPF implementations in Table
\ref{tab:results} of Section \ref{sec:results}.
However, a recent result by \cite{Jones2013} allows one to enact a
QFT using only Clifford gates and a Toffoli gate in $O(\log^2 n)$ expected depth.
This would allow us to
greatly improve the constants in our circuit resource upper bounds in Section \ref{sec:modexp} by combining a QFT with parallel multiplication similar to
the approach described in \cite{VanMeter2005,Cleve2000}.

We also note that recent results by Browne, Kashefi, and Perdrix (BKP) connect the power of
measurement-based quantum computing to the quantum circuit model augmented with
unbounded fanout \cite{Browne2009}. Their model, which we adapt and call
\textsc{CCNTC}, uses the classical controller mentioned in Section \ref{subsec:fanout}.
Using results by H{\o}yer and {\v S}palek \cite{Hoyer2002} that
unbounded quantum fanout would allow for a constant-depth factoring algorithm,
they conclude that a probabilistic polytime classical machine with access
to a constant-depth one-way quantum computer would also be able to factor
efficiently.

\section{The Constant-Depth Carry-Save Technique}
\label{sec:csa}

Our 2D factoring approach rests on the central technique of the constant-depth
carry-save adder (CSA) \cite{Gossett1998}, which converts the sum of three
numbers $a$, $b$, and $c$, to the sum of two numbers $u$ and $v$:
$a+b+c = u+v$. The explanation of this technique and how it achieves constant depth requires the following definitions.

A \emph{conventional number} $x$ can be represented in $n$ bits as
$x = \sum_{i=0}^{n-1} 2^i x_i$,
where $x_i \in \{0,1\}$ denotes the $i$-th bit of $x$, which we call
an $i$-bit and has significance $2^i$, and the $0$-th bit is the low-order bit.\footnote{It will be clear from the context whether we mean an
$i$-bit, which has significance $2^i$, or an $i$-bit number.}
Equivalently, $x$ can be represented as a (non-unique)
sum of two smaller, $(n-1)$-bit, conventional numbers, $u$ and $v$.
We say $(u+v)$ is a \emph{carry-save encoded}, or CSE, number.
The CSE representation of an $n$-bit conventional number
consists of $2n-2$ individual
bits where $v_0$ is always $0$ by convention.

Consider a CSA operating on three bits instead of three numbers; 
then a CSA converts the sum of three
$i$-bits into the sum of an $i$-bit (the \emph{sum} bit) and an $(i+1)$-bit
(the \emph{carry} bit):
$a_i+b_i+c_i = u_i+v_{i+1}$.
By convention, the bit $u_i$ is the parity of the input bits
($u_i = a_i \oplus b_i \oplus c_i$) and
the bit $v_{i+1}$ is the majority of $\{a_i, b_i, c_i\}$.
Figure \ref{fig:csa-encoding} gives a concrete example, where
$(u+v)$ has $2n-2 = 8$ bits, not counting $v_0$.

It will also be useful to refer to a subset of the bits in a conventional
number using subscripts to indicate a range of indices:
\begin{equation}
x_{(j,k)} \equiv \sum_{i=j}^k 2^ix_i \qquad
x_{(i)} \equiv x_{(i,i)} = 2^ix_i.
\end{equation}
Using this notation, the following identity holds:
\begin{equation}
x_{(j,k)} = x_{(j,\ell)} + x_{(\ell+1,k)}, \qquad \text{ for all } j \le \ell < k.
\end{equation}
We can express the relationship between the bits of $x$ and $(u+v)$ as follows:
\begin{equation}
x = x_{(0,n-1)} \equiv u+v = u_{(0,n-2)} + v_{(1,n-1)}.
\end{equation}
Finally, we denote arithmetic modulo $m$ with square brackets.

\begin{equation}
x_{(j,k)} \bmod m = x_{(j,k)}[m]
\end{equation}

\begin{center}
\begin{figure*}[tb!]
\begin{displaymath}
x = 30 = u+v = 8 + 22 = \left\{
\begin{array}{ccccc}
    & u_3 & u_2 & u_1 & u_0 \\
v_4 & v_3 & v_2 & v_1 &    \\
\hline
x_4 & x_3 & x_2 & x_1 & x_0
\end{array}
\right\}
=
\left\{
\begin{array}{ccccc}
    & 1 & 0 & 0 & 0 \\
  1 & 0 & 1 & 1 &   \\
\hline
1 & 1 & 1 & 1 & 0
\end{array}
\right\}
\end{displaymath}
\caption{An example of carry-save encoding for the 5-bit conventional number 30.}
\label{fig:csa-encoding}
\end{figure*}
\end{center}

\begin{figure}[tb!]
\begin{center}
\begin{displaymath}
\centerline{
\Qcircuit @C=2em @R=2em {
\lstick{\ket{0}}   & \qw      & \qw & \qw                        & \qw & \qw                        & \targfix  & \qw & \qw_{\ket{a_i \wedge (b_i \oplus c_i)}} & \targfix  & \qw       & \qw       & \qw_{\ket{(b_i \wedge c_i) \oplus a_i \wedge (b_i \oplus c_i)}} & \qw & \qswap      & \qswap      & \qw & \rstick{\ket{u_i}} \\
\lstick{\ket{a_i}} & \qw      & \qw & \qw                        & \qw & \qw                        & \ctrl{-1} & \qw & \qw                                     & \qw       & \targfix  & \qw       & \qw_{\ket{a_i \oplus b_i \oplus c_i}}                           & \qw & \qw \qwx    & \qswap \qwx & \qw & \rstick{\ket{0}} \\
\lstick{\ket{b_i}} & \ctrl{1} & \qw & \targfix                   & \qw & \qw_{\ket{b_i \oplus c_i}} & \ctrl{-1} & \qw & \qw                                     & \qw       & \ctrl{-1} & \targfix  & \ctrl{1}                                                        & \qw & \qw \qwx    & \qw         & \qw & \rstick{\ket{b_i}} \\
\lstick{\ket{c_i}} & \ctrl{1} & \qw & \ctrl{-1}                  & \qw & \qw                        & \qw       & \qw & \qw                                     & \qw       & \qw       & \ctrl{-1} & \ctrl{1}                                                        & \qw & \qw \qwx    & \qw         & \qw & \rstick{\ket{c_i}} \\
\lstick{\ket{0}}   & \targfix & \qw & \qw_{\ket{b_i \wedge c_i}} & \qw & \qw                        & \qw       & \qw & \qw                                     & \ctrl{-4} & \qw       & \qw       & \targfix                                                        & \qw & \qswap \qwx & \qw         & \qw & \rstick{\ket{v_{i+1}}}}
}
\end{displaymath}
\caption{Carry-save adder circuit for a single bit position $i$: $a_i+b_i+c_i = u_i + v_{i+1}$.}
\label{fig:csa-circuit}
\end{center}\end{figure}

\begin{figure}
\begin{center}
\begin{displaymath}
\begin{tabular}{p{0.5in} m{0.1in} p{2in}}

\Qcircuit @C=1em @R=2.2em { 
	& \qw & \ctrl{1} & \qw & \qw \\
	& \qw & \ctrl{1} & \qw & \qw \\
	& \qw & \targfix & \qw & \qw
}

&
\qquad
=
\qquad
&

\Qcircuit @C=1em @R=.7em { 
	& \gate{T^{\dagger}} & \qw & \targfix  & \qw & \gate{T} & \qw & \targfix  & \qw & \gate{T^{\dagger}} & \qw & \targfix  & \qw & \gate{T}           & \qw & \targfix  & \qw & \qw \\ 
	& \gate{T^{\dagger}} & \qw & \qw       & \qw & \ctrl{1} & \qw & \ctrl{-1} & \qw & \ctrl{1}           & \qw & \qw       & \qw & \qw                & \qw & \ctrl{-1} & \qw & \qw \\
	& \gate{H}           & \qw & \ctrl{-2} & \qw & \targfix & \qw & \gate{T}  & \qw & \targfix           & \qw & \ctrl{-2} & \qw & \gate{T^{\dagger}} & \qw & \gate{H}  & \qw & \qw
}
\end{tabular}
\end{displaymath}
\caption{The depth-efficient Toffoli gate decomposition from \cite{Amy2012}.}
\label{fig:toffoli}
\end{center}
\end{figure}

\begin{table}
\begin{displaymath}
\begin{tabular}{|c|c|c|c|}
\hline
\text{Circuit Name} & \text{Depth} & \text{Size} & \text{Width} \\
\hline
\text{Toffoli gate from \cite{Amy2012}} and Figure \ref{fig:toffoli} & 8 & 15 & 3 \\
\hline
\text{Single-bit } 3\text{-to-}2 \text{ adder from Figure \ref{fig:csa-circuit}} & 33 & 55 & 5 \\
\hline
\end{tabular}
\end{displaymath}
\centerline{}
\caption{Circuit resources for Toffoli and single-bit addition.}
\label{tab:csa-tile-resources}
\end{table}

Figure \ref{fig:csa-circuit} gives a circuit description of carry-save addition (CSA) for a single bit position $i$.
The resources for this circuit are given in Table \ref{tab:csa-tile-resources}, using
the resources for the Toffoli gate (in the same table) based on
\cite{Amy2012}. We note here
that a more efficient decomposition for the Toffoli is possible using a
distillation approach described in \cite{Jones2013a}.

\begin{figure}[b!]
\begin{center}
\includegraphics[width=3in]{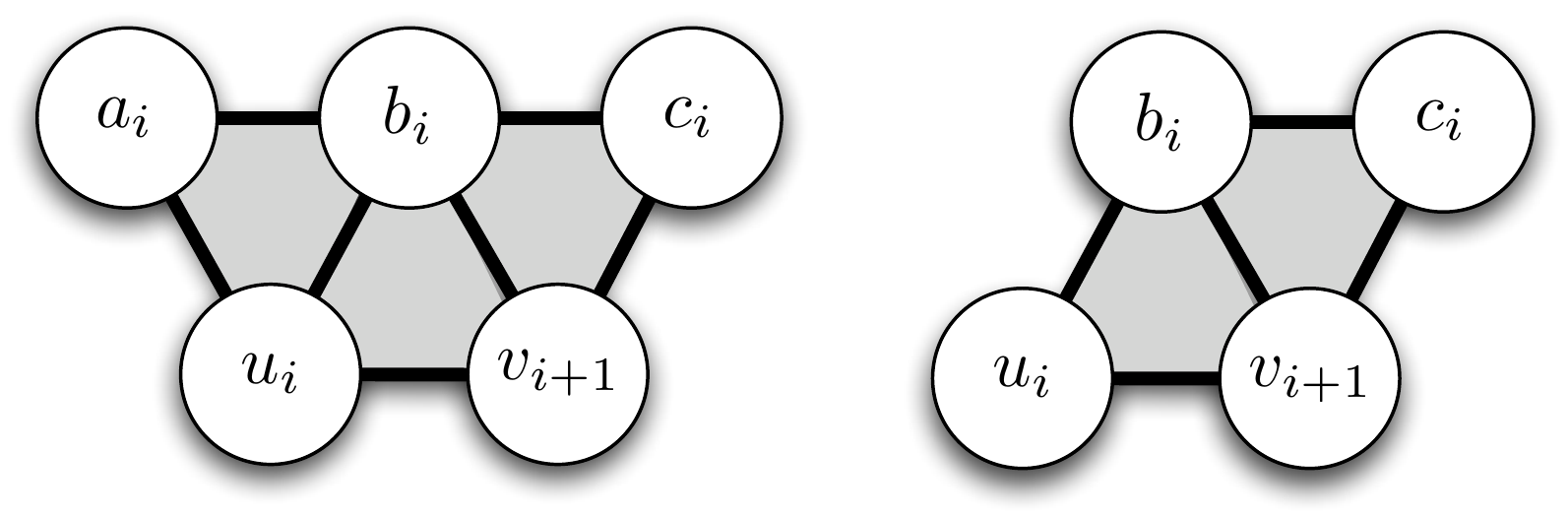}
\end{center}
\caption{The carry-save adder (CSA), or 3-2 adder, and carry-save 2-2 adder.}
\label{fig:csa-3-2}
\end{figure}

We must lay out the circuit to satisfy a 2D NTC model.
The Toffoli gate decomposition in \cite{Amy2012}, duplicated in
Figure \ref{fig:toffoli}, requires two control
qubits and a single target qubit to be
mutually connected to each other. Given this constraint, and the
interaction of the CNOTs in Figure \ref{fig:csa-circuit}, we can
rearrange these qubits on a 2D planar grid and obtain the layout shown
in Figure \ref{fig:csa-3-2}, which satisfies our 2D NTC model.
Qubits $\ket a_i$, $\ket b_i$, and $\ket c_i$ reside at the top of Figure~\ref{fig:csa-3-2}, while qubits $\ket{u_i}$ and $\ket{v_{i+1}}$ are initialized to $\ket 0$.
Upon completion of the circuit, qubit $\ket{a_i}$ is in state $\ket 0$, as seen from the output in Figure~\ref{fig:csa-circuit}. 
Note that this construction uses more gates and one more ancilla than the equivalent
quantum full adder circuit in Figure 5 of \cite{Gossett1998}. However this
is necessary in order to meet our architectural constraints and does not change the
asymptotic results.
Also in Figure \ref{fig:csa-3-2}
is a variation called a 2-2 adder, which simply re-encodes two $i$-bits
into an $i$-bit and an $(i+1)$-bit. The 2-2 adder uses at most the resources
of a 3-2 adder, so we can count it as such in our calculations.
It will be useful in the next section.

At the level of numbers, the sum of three $n$-bit numbers can be converted into
the sum of two $n$-bit numbers by applying a \emph{CSA layer} of
$n$ parallel, single-bit
CSA circuits (Fig.~\ref{fig:csa-circuit}). Since each CSA operates in constant depth, the entire layer also
operates in constant depth, and we have achieved (non-modular) addition.
%
Each single addition of three $n$-bit numbers requires $O(n)$ circuit width.

\section{Quantum Modular Addition}
\label{sec:csa-mod-add}

To perform addition of two numbers $a$ and $b$ modulo $m$,
we consider the variant problem of modular addition of three numbers to
two numbers:
%
Given three $n$-bit input numbers $a$, $b$, and $c$, and an $n$-bit modulus $m$,
compute
$(u+v) = (a+b+c)[m]$,
where $(u+v)$ is a CSE number.

In this section, we provide an alternate, pedagogical explanation of
Gossett's modular reduction \cite{Gossett1998}. Later, we contribute a mapping of this adder
to a 2D architecture,
using unbounded fanout to maintain constant depth for adding back
modular residues. This last step is absent in Gossett's original approach.

To start, we will demonstrate the basic method of modular addition and reduction
on an $n$-bit conventional number. In general, adding two $n$-bit conventional
numbers will produce an overflow bit of significance $2^n$, which we can truncate as long as
we add back its modular residue $2^n \bmod m$. How can we guarantee that we won't
generate another overflow bit by adding back the modular residue? It turns out
we can accomplish this by allowing
a slightly larger input and output number ($n+1$ bits in this case), truncating
multiple overflow bits, and adding back their $n$-bit modular residues.

For two $(n+1)$-bit conventional numbers $x$ and $y$,
we truncate the three high-order bits of their sum $z_{n-1,n+3}$
and
add back their modular residue $x_{(n-1,n)}[m]$:
\begin{eqnarray}
x + y \bmod m &=& z_{(0,n+1)}[m] \nonumber \\
&=& z_{(0,n-2)} + z_{(n-1,n+1)}[m].
\end{eqnarray}
Since both the truncated number $z_{(0,n-2)}$ and the modular residue
are $n$-bit numbers, their sum is an $(n+1)$-bit number as desired, equivalent
to $x[m]$.

Now we must do the same modular reduction on a CSE number $(u+v)$,
which in this case represents an $(n+2)$-bit conventional number and has
$2n+3$ bits.
%
%
First, we truncate the three high-order bits ($v_{n}, u_{n-1}, v_{n-1}$)
of $(u+v)$, yielding an $n$-bit
conventional number with a CSE representation of $2n$ bits:
$\{u_0, u_1, \ldots, u_{n-1}\} \cup \{v_1, v_2, \ldots, v_{n-1}\}$.
Then we add back the three modular residues
$(v_{(n+1)}[m], u_{(n)}[m], v_{(n)}[m])$, and we are guaranteed not to
generate additional overflow bits (of significance $2^{n}$ or higher). This equivalence
is shown in Equation \ref{eqn:mod-reduce}.
\begin{eqnarray}
(u+v)[m] &=& \left(u_{(0,n+1)} + v_{(1,n+2)}\right)[m] \nonumber \\
 &=& u_{(0,n)} +
     v_{(1,n)} + \nonumber \\
 & & u_{(n+1)}[m] +
     v_{(n+1)}[m] + v_{(n+2)}[m]
\label{eqn:mod-reduce}
\end{eqnarray}

\begin{lemma}[Modular Reduction in Constant Depth]
The modular addition of three $n$-bit numbers to two $n$-bit numbers can be
accomplished
in constant depth with $O(n)$ width in \textsc{2D CCNTC}.
\end{lemma}

\vspace*{12pt}
\noindent
{\bf Proof:}
Our goal is to show how to perform modular addition while keeping our numbers
of a fixed size by treating overflow bits correctly.
We map the proof of \cite{Gossett1998} to \textsc{2D CCNTC} and show that
we meet our required depth and width.
First, we enlarge our registers to allow the addition of $(n+2)$-bit numbers,
while keeping our modulus of size $n$ bits.
(In Gossett's original approach, he takes the equivalent step of restricting
the modulus to be of size $(n-2)$ bits.) We accomplish the modular addition
by first performing a layer of non-modular addition, truncating the three high-order
overflow bits, and then adding back modular residues controlled on these
bits in three successive layers, where we are guaranteed that no additional
overflow bits are generated in each layer.
This is illustrated for a $3$-bit modulus and $5$-bit registers
in Figure \ref{fig:csa-proof}.

\begin{center}
\begin{figure*}[h!tb]
\begin{displaymath}
\renewcommand\arraystretch{1.5}
\begin{array}{ccccccll}
        & a_4 & a_3 & a_2 & a_1 & a_0 & 5\text{-bit input number } a &\\
        & b_4 & b_3 & b_2 & b_1 & b_0 & 5\text{-bit input number } b & \\
        & c_4 & c_3 & c_2 & c_1 & c_0 & 5\text{-bit input number } c & \text{[Layer 1]}\\
\hline
        & u_4 & u_3 & u_2 & u_1 & u_0 & \text{truncate } u_{4} & \\
    v_5 & v_4 & v_3 & v_2 & v_1 &     & \text{truncate } v_{4},v_{5} & \\
        &     &     & c^{v_4}_2 & c^{v_4}_1 & c^{v_4}_0 & \text{add back } 2^4 \bmod m \text{ controlled on } v_4 & \text{[Layer 2]}\\
\hline
        &      & u'_3 & u'_2 & u'_1 & u'_0 & & \\
        & v'_4 & v'_3 & v'_2 & v'_1 &      & & \\
        &      &    & c^{u_4}_2 & c^{u_4}_1 & c^{u_4}_0  & \text{add back } 2^4 \bmod m \text{ controlled on } u_4 & \text{[Layer 3]}\\
\hline
        & u''_4 & u''_3 & u''_2 & u''_1 & u''_0 & \text{the bit } u''_4 \text{ is the same as } v'_4 & \\
        & v''_4 & v''_3 & v''_2 & v''_1 &       &  & \\
        &       &    & c^{v_5}_2 & c^{v_5}_1 & c^{v_5}_0 & \text{add back } 2^5 \bmod m \text{ controlled on } v_5 & \text{[Layer 4]}\\
\hline
        & u'''_4 & u'''_3 & u'''_2 & u'''_1 & u'''_0 & \text{ Final CSE output with } 5 \text{ bits} &\\
        & v'''_4 & v'''_3 & v'''_2 & v'''_1 &        & \text{ Final CSE output with } 5 \text{ bits} & \\
\end{array}
\end{displaymath}
\caption{A schematic proof of Gossett's constant-depth modular reduction for $n=3$.}
\label{fig:csa-proof}
\end{figure*}
\end{center}

We use the following notation.
The non-modular sum of the first layer is $u$ and $v$.
The CSE output of the first modular reduction layer
is $u'$ and $v'$, and the modular residue is
written as $c^{v_{n+1}}$ to mean the precomputed value $2^{n+1} \bmod m$
controlled on $v_{n+1}$.
The CSE output of the second modular reduction layer
is $u''$ and $v''$, and the modular residue is written as
$c^{u_{n+1}}$ to mean the precomputed value $2^{n+1} \bmod m$
controlled on $u_{n+1}$.
The CSE output of the third and final modular reduction layer
is $u'''$ and $v'''$, and the modular residue is written as
$c^{v_{n+2}}$ to mean the precomputed value $2^{n+2} \bmod m$
controlled on $v_{n+2}$.

We show that no layer generates an overflow $(n+2)$-bit, namely in the
$v$ component of any CSE output. (The $u$ component will never exceed the
size of the input numbers.) First, we know that no $v'_{n+2}$ bit
is generated after the first modular reduction layer, because we have
truncated away all $(n+1)$-bits. Second, we know that no $v''_{n+2}$ bit is
generated because we only have one $(n+1)$-bit to add, $v'_{n+1}$.
Finally, we need to show that $v'''_{n+2} = 0$ in the third modular reduction
layer. 

Since $u'_{(n)} + v'_{(n+1)} =
u_{(n)} + v_{(n)} \le 2^{n+1}$, the bits $u'_n$ and $v'_{n+1}$ cannot both be $1$.
But $u''_{n+1} = v'_{n+1}$ and $v''_{n+1} = u'_n\land v'_n$, so $u''_{n+1}$ and
$v''_{n+1}$ cannot both be $1$, and hence $v'''_{n+2} = 0$.
Everywhere
we use the fact that the modular residues are restricted to $n$ bits.
Therefore, the modular sum is computed as the sum of two $(n+2)$-bit numbers
with no overflows in constant-depth.
$\square$\,

As a side note, we can perform modular reduction in one layer instead of
three by decoding the three overflow bits into one of seven different
modular residues. This can also be done in constant depth, and in this case
we only need to enlarge all our registers to $(n+1)$ bits instead of $(n+2)$
as in the proof above. We omit the proof for brevity.

In the following two subsections, we give a concrete example to illustrate
the modular addition circuit as well as a numerical upper bound for the
general circuit resources.

\subsection{A Concrete Example of Modular Addition}
\label{subsec:concrete}

\begin{center}
\begin{figure*}[h!bt]
\centerline{
\includegraphics[width=6.5in]{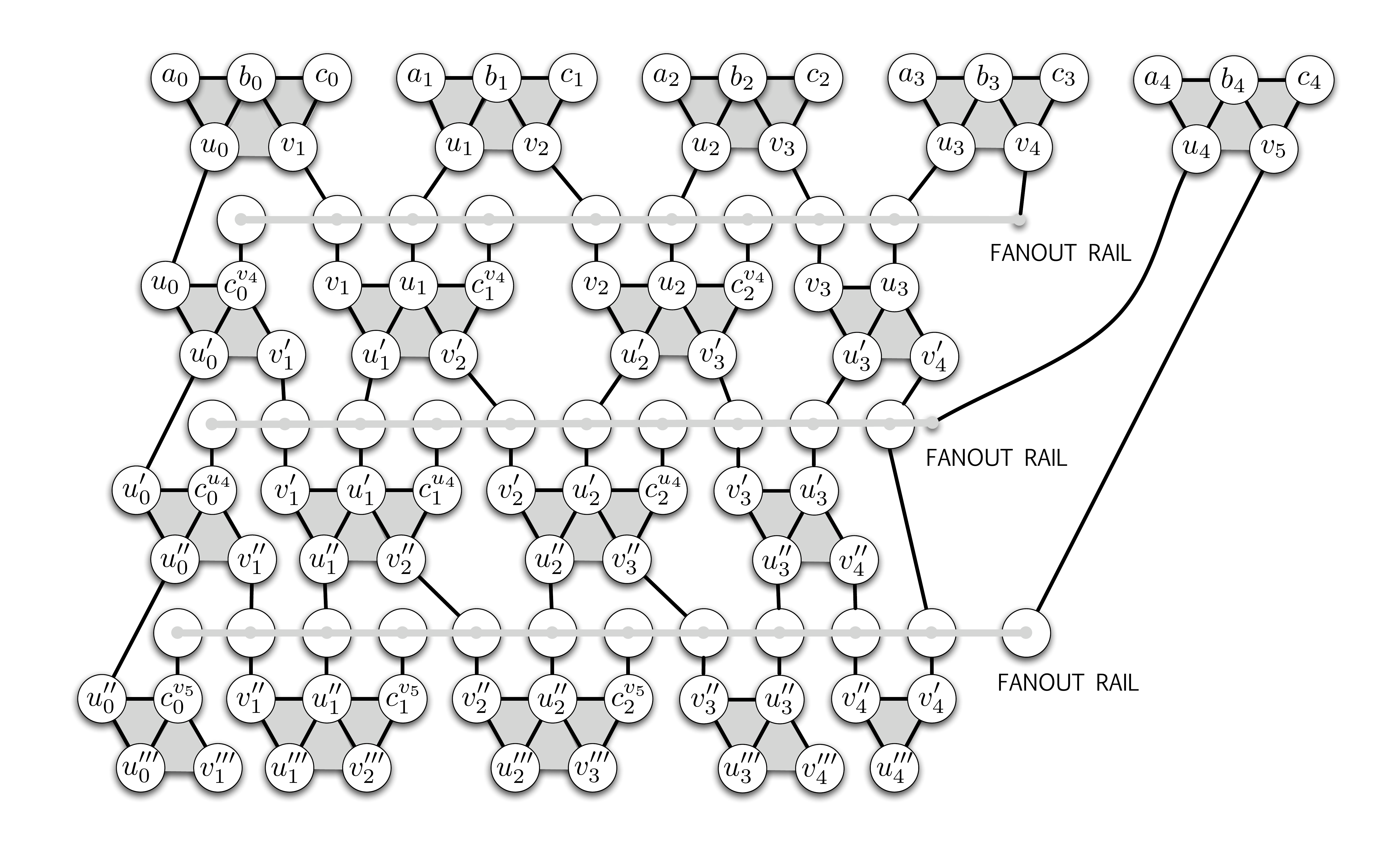}
}
\caption{Addition and three rounds of modular reduction for a 3-bit
modulus.}
\label{fig:csa-add-4}
\end{figure*}
\end{center}

A \textsc{2D CCNTC} circuit for modular addition of $5$-bit numbers using
four layers of parallel CSA's is shown graphically in Figure \ref{fig:csa-add-4}
which corresponds directly to the schematic proof in Figure \ref{fig:csa-proof}.
Note that in Figure \ref{fig:csa-add-4}, the least significant qubits are
on the left, and in Figure \ref{fig:csa-proof}, the least significant qubits are
on the right.
Figure \ref{fig:csa-add-4} also represents the approximate
physical layout of the qubits as they would look if this
circuit were to be fabricated.
Here, we convert the sum of three
$5$-bit integers into the modular sum of two $5$-bit integers, with a
$3$-bit modulus $m$.
In the first layer,
we perform 4 CSA's in parallel on the input numbers ($a,b,c$) and produce the
output numbers ($u, v$).

As described above, we truncate
the three high-order bits during the initial CSA round
(bits $u_4, v_4, v_5$) to retain a $4$-bit number.
Each of these bits serves as a control for adding its modular residue to
a running total. We can classically precompute $2^4[m]$ for the two
additions controlled on $u_4$ and $v_4$ and
$2^5[m]$ for the addition controlled on $v_5$.

In Layer 2,
we use a constant-depth fanout rail (see Figure \ref{fig:cdf}) to
distribute the control bit $v_4$ to its modular residue, which we denote as
$\ket{c^{v_4}} \equiv \ket{2^4[m]\cdot v_4}$.
The register $c^{v_4}$ has $n$ bits, which we add to the CSE results of layer 1.
The results $u_i$ and $v_{i+1}$ are teleported into layer 3. The exception is
$v'_4$ which is teleported into layer 4, since there are no other $4$-bits
to which it can be added. Wherever there are only
two bits of the same significance, we use the 2-2 adder from
Section \ref{sec:csa}.

Layer 3
operates similarly to layer 2, except that the modular residue is controlled on
$u_4$:
$\ket{c^{u_4}} \equiv \ket{2^4[m] \cdot u_4}$.
The register $c^{u_4}$ has $3$ bits, which we
add to the CSE results of layer 2, where $u'_i$ and $v'_{i+1}$ are teleported
forward into layer 4.

Layer 4
is similar to layers 2 and 3, with the modular residue controlled on $v_5$:
$\ket{c^{v_5}} \equiv \ket{2^5[m] \cdot v_5}$.
The register $c^{v_5}$ has $3$ bits, which we
add to the CSE results of layer 3.
There is no overflow bit $v'''_5$, and no carry bit from $v''_4$ and $v'_4$
as argued in the proof of Lemma 1.
The final modular sum $(a+b+c)[m]$ is $u'''+v'''$.

\begin{center}
\begin{figure*}[h!bt]
\centerline{
\includegraphics[width=1.5in]{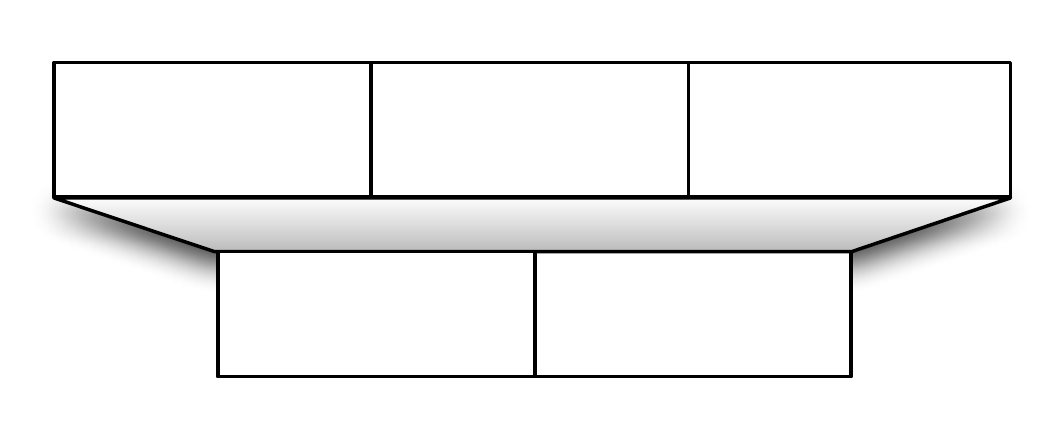}
}
\caption{Symbol for an $n$-bit 3-to-2 modular adder, also called a CSA tile.}
\label{fig:csa-tile-symbol}
\end{figure*}
\end{center}

The general circuit for adding three $n$-qubit quantum integers to
two $n$-qubit quantum integers is called a \emph{CSA tile}. Each CSA tile in our architecture 
corresponds to its own module, and it will be represented by the symbol in 
Figure \ref{fig:csa-tile-symbol} for the rest of this paper. We call this
an $n$-bit modular adder, even though it accepts $(n+2)$-bit inputs, because
the size of the modulus is still $n$ bits.

\subsection{Quantum Circuit Resources for Modular Addition}

We now calculate numerical upper bounds for the circuit resources of
the $n$-bit $3$-to-$2$ modular adder described in the previous section.
There are four layers of non-modular $n'$-bit $3$-to-$2$ adders, each of which
consists of $n'$ parallel single-bit adders whose
resources are detailed in Table \ref{tab:csa-tile-resources}. For factoring
an $n$-bit modulus, we have $n'=n+2$ in the first and fourth layers
and $n'=n+1$ in the second and third layers.

After each of the first three layers, we must move the output qubits
across the fanout rail to be the inputs of the next layer. We use
two swap gates, which have a depth and size of $6$ CNOTs each, since
the depth of teleportation is only more efficient for moving more than
two qubits. The control bit for each modular residue needs to be
teleported $0$, $4$, and $7$ qubits respectively according to the
diagram in Figure \ref{fig:csa-add-4}, before being fanned out $n$
times along the fanout rails, where the fanned out copies will end up
in the correct position to be added as inputs.


The resources for the $n$-bit $3$-to-$2$ modular adder depicted in Figure
\ref{fig:csa-add-4} are given below.
The formulae reflect the resources needed for both computing the output
in the forward direction (including creating an entangled fanned-out state
controlled on overflow qubits)
and also uncomputing ancillae in the backward
direction (including disentangling previous fanned-out copies).

The circuit depth is $O(1)$:

\begin{equation}
374\text{.}
\end{equation}

The circuit size is $O(n)$:

\begin{equation}
551n + 757\text{.}
\end{equation}

The circuit width is $O(n)$:

\begin{equation}
33n + 47\text{.}
\end{equation}

\section{Quantum Modular Multiplication}
\label{sec:csa-mod-mult}

We can build upon our carry-save adder to implement quantum modular
multiplication in logarithmic depth. We start with a completely classical
problem to illustrate the principle of multiplication by repeated addition.
Then we consider modular multiplication of two quantum integers in a serial
and a parallel fashion in Section
\ref{subsec:csa-mod-mult-qq}. Both of these problems use as subroutines
\emph{partial product creation}, which we define and solve
 in Section \ref{subsec:ppc} and
 \emph{modular multiple addition}, which we define and solve
in Section \ref{subsec:mma}.

First we consider a completely classical problem:
given three $n$-bit classical numbers $a$, $b$, and $m$,
compute $c = ab \bmod m$, where $c$ is allowed to be in CSE.

We only have to add shifted
multiples of $a$ to itself, ``controlled'' on the bits of $b$. There are
$n$ shifted multiples of $a$, let's call them $z^{(i)}$, one for every bit of $b$:
$z^{(i)} = 2^i a b_i \bmod m$.
We can parallelize the addition of $n$ numbers in a logarithmic depth
binary tree to get a total depth of $O(\log n)$.

\subsection{Modular Multiplication of Two Quantum Integers}
\label{subsec:csa-mod-mult-qq}

We now consider the problem of multiplying a classical number controlled
on a quantum bit with a
\emph{quantum integer}\footnote{In this paper, an $n$-qubit 
quantum integer is a
general superposition of up to $2^n$ classical integers. As a special case,
a classical number controlled on a single qubit is a superposition of
$2$ classical integers.},
which is a
quantum superposition of classical numbers:

\begin{quote}
Given an $n$-qubit quantum integer $\ket{x}$, a control qubit $\ket{p}$,
and two $n$-bit classical numbers $a$
and $m$,
compute $\ket{c} = \ket{xa[m]}$, where $c$ is allowed to be in CSE.
\end{quote}

This problem occurs naturally in modular exponentiation (described in
the next section) and can be considered \emph{serial multiplication},
in that $t$ quantum integers are multiplied in series to a single
quantum register. This is used in serial QPF as mentioned in
Section \ref{sec:related}.

We first create $n$ quantum integers $\ket{z^{(i)}}$,
which are shifted multiples of the classical number $a$ controlled on the bits
of $x$:
$\ket{z^{(i)}} \equiv \ket{2^i a[m] \cdot x_i }$.
These are typically called \emph{partial products} in a classical multiplier.
How do we create these numbers, and what is the depth of the procedure?
First, note that $\ket{2^i a[m]}$ is a classical number, so we can
precompute them classically and prepare them in parallel using single-qubit
operations
on $n$ registers, each consisting of $n$ ancillae qubits. Each $n$-qubit
register will hold a future $\ket{z^{(i)}}$ value.
We then fan out each of the
$n$ bits of $x$, $n$ times each, using an unbounded fanout operation so that
$n$ copies of each bit $\ket{x_i}$ are next to register $\ket{z^{(i)}}$.
This takes a total of $O(n^2)$ parallel CNOT operations.
We then entangle each $\ket{z^{(i)}}$ with the corresponding $x_i$.
After this, we interleave these numbers into groups of three using
constant-depth teleportation. This reduces to the task of modular
multiple addition in order to add these numbers down to a single
(CSE) number modulo $m$, which is described in Section \ref{subsec:mma}.


Finally, we tackle the most interesting problem:
\begin{quote}
Given two $n$-qubit quantum integers $\ket{x}$ and
$\ket{y}$ and an $n$-bit classical number
$m$,
compute $\ket{c} = \ket{xy \bmod m}$,
where $\ket{c}$ is allowed to be in CSE.
\end{quote}

This can be considered \emph{parallel multiplication} and is responsible
for our logarithmic speedup in modular exponentiation and parallel QPF.

Instead of creating $n$ quantum integers $\ket{z^{(i)}}$, we must create
up to $n^2$ numbers
$\ket{z^{i,j}}$ for all possible pairs of quantum bits $x_i$ and $y_j$,
$i,j \in \{0,\ldots,n-1\}$:
$\ket{z^{i,j}} \equiv \ket{2^i2^j[m]\cdot x_i \cdot y_j}$.
We create these numbers using a similar procedure to the previous problem.
Adding $n^2$ quantum integers of $n$ qubits each takes depth
$O(\log(n^2))$, which is still $O(\log n)$.
Creating $n^2\times n$-bit quantum integers takes width $O(n^3)$.
Numerical constants are given for these resource estimates in
Section \ref{subsec:mod-mult-resources} for the entire modular multiplier.

Here is an outline of our modular multiplier construction, combining the
two halves of partial product creation (Section \ref{subsec:ppc}) and
modular multiple addition (Section \ref{subsec:mma}).

\begin{enumerate}
\item Initially, the inputs consist of the CSE quantum integers $x$ and $y$,
each with $2n+3$ bits, sitting on adjacent edges of a square lattice that has
sides of length $3(2n+3)$ qubits.
\item For each of $\lceil \log_2 (2n+3) \rceil$ rounds:
\begin{enumerate}
\item Of the existing $\{x_i\}$ and $\{y_j\}$ bits, apply a CNOT to create an
entangled copy in an adjacent qubit.
\item Teleport this new copy halfway between its current location and the
new copy.
\item At every site where an $\ket{x_i}$ and an $\ket{y_j}$ meet,
apply a Toffoli gate to create $\ket{x_i \cdot y_j}$.
\item Teleport $\ket{x_i \cdot y_j}$ to the correct $z$-site module.
\end{enumerate}
\item Within each $z$-site module, fanout $\ket{x_i \cdot y_j}$ up to $n$
times, corresponding to each $1$ in the modular residue $2^i 2^j \bmod m$,
to create the $n$-qubit quantum integer $\ket{z^{(i,j)}}$.
\item For each triplet of $z$-site modules, teleport the quantum integers
$\ket{z^{(i,j)}}$ to a CSA tile module, interleaving the three numbers so that
bits of the same significance are adjacent. This concludes partial product
creation (Section \ref{subsec:ppc}).
\item Perform modular multiple addition (described in Section \ref{subsec:mma})
on $t'$ $n$-qubit quantum integers down to 2 $n$-qubit quantum integers (one CSE number).
\item Uncompute all the previous steps to restore ancillae to $\ket{0}$.
\end{enumerate}
\subsection{Partial Product Creation}
\label{subsec:ppc}

This subroutine describes the procedure of creating $t'=O(n^2)$ partial products of
the CSE quantum integers $x$ and $y$, each with $2n+3$ bits each. We will now
discuss only the case of parallel multiplication. Although we
will not provide an explicit circuit for this subroutine, we will outline
our particular construction and give a numerical upper bound on the
resources required.

First, we need to generate the product bits
$\ket{x_i\cdot y_j}$ for all possible $(2n+3)^2$ pairs of $\ket{x_i}$ and
$\ket{y_j}$.
A particular product bit $\ket{x_i \cdot y_j}$
controls a particular classical number, the
$n$-bit modular residue $2^i 2^j [m]$, to form the partial product
$\ket{z^{(i,j)}}$ defined
in the previous section. However, some of these partial products
consist of only a single qubit, if $2^i 2^j < 2^n$, which is the minimum
value for an $n$-bit modulus $m$. There are at least $2n^2 - 2n + 1$
such single-bit partial products, which can be grouped into at most
$(2n+3)\times n$-bit numbers. Of the $(2n+3)^2$ possible partial products,
this leaves the number of remaining $n$-bit partial products as at most
$2n^2 + 14n +8$. Therefore we have a maximum number of $n$-bit
partial products, which we will simply refer to as $t'$ from now on.

\begin{equation}
t'=2n^2+16n+11
\label{eqn:tprime}
\end{equation}

The creation of the product bits $\ket{x_i \cdot y_j}$ occurs on a
square lattice of $(3(2n+3))^2$ qubits, with the numbers $\ket{x_i}$ and
$\ket{y_j}$ located on adjacent edges. The factor of $3$ in the size of the lattice
allows the $\ket{x_i}$ and $\ket{y_j}$ bits move past each other.
The $\ket{x_i}$ bits are teleported along an axis that is perpendicular to
the teleportation axis for the $\ket{y_j}$ bits, and vice versa.
Product bit creation, and this square lattice, comprise a single module.
In $\lceil \log_2 (2n+3) \rceil$
rounds, these bits are copied via a CNOT and teleported to the middle of
a recursively halved interval of the grid. The copied bits $\ket{x_i}$ and
$\ket{y_j}$
first form $1$ line, then $3$ lines, then $7$ lines, and so forth,
intersecting at $1$ site, then $9$ sites, then $49$ sites, and so forth.
There are $\lceil \log_2 (2n+3) \rceil$ such rounds.

At each intersection, a Toffoli gate is used to create $\ket{x_i \cdot y_j}$
from the given $\ket{x_i}$ and $\ket{y_j}$. These product bits are then
teleported away from this qubit, out of this product bit module, to different
modules where the $\ket{z^{(i,j)}}$ numbers are later generated,
called $z$-sites. There are $t'$ $z$-site modules which each contain 
an $n$-qubit quantum integer. Any
round of partial product generation will produce at most as many product
bits $x_i \cdot y_j$ as in the last round, which is half the total number
of $(2n+3)^2$.


We now present the resources for partial product creation, the first half of
a modular multiplier, including the reverse computation.

The circuit depth is $O(\log n)$:

\begin{equation}
D_{PPC} = 32\log_2 n + 150\text{.}
\end{equation}

The module depth is $O(1)$:

\begin{equation}
\overline{D}_{PPC} = 8\text{.}
\end{equation}

The circuit size is $O(n\log^2 n)$:

\begin{eqnarray}
S_{PPC} & = & (6n + 9)\log_2 n +\\
        &   & (26n^3 + 232n^2 + 224n + 159)\text{.}
\end{eqnarray}

The module size is $O(n^2)$:

\begin{eqnarray}
\overline{S}_{PPC} = 6n^2 + 26n + 19\text{.}
\end{eqnarray}

The circuit width is $O(n^3)$:

\begin{eqnarray}
W_{PPC} = 6n^3 + 48n^2 - 8n + 1\text{.}
\end{eqnarray}

The module width is $O(n^2)$:

\begin{eqnarray}
\overline{W}_{PPC} = 2n^2 + 14n + 9\text{.}
\end{eqnarray}

\subsection{Modular Multiple Addition}
\label{subsec:mma}

As a subroutine to modular multiplication, we define the operation of
repeatedly adding multiple numbers down to a single CSE number, called
\emph{modular multiple addition}.

The modular multiple addition circuit generically adds down $t'\times n$-bit
conventional numbers to an $n$-bit CSE number:
\begin{equation}
z^{(1)} + z^{(2)} + \ldots z^{(t')} \equiv (u+v)[m].
\end{equation}
It does not matter how the
$t'$ numbers are generated, as long as they are divided into groups of three
and have their bits interleaved to be the inputs of a CSA tile.
From the previous section, serial multiplication results in
$t' \le n$ and parallel multiplication results in $t' \le n^2$. Each CSA tile
is contained in its own module. These modules are arranged in layers within
a logarithmic depth binary tree, where 
the first layer contains $\lceil t'/3 \rceil$ modules. A modular addition
occurs in all the modules of the first layer in parallel. The outputs from this
first layer are then teleported to be the inputs of the next layer of modules,
which have at most two-thirds as many modules. This continues until the
tree terminates in a single module, whose output is a CSE number $u+v$ which
represents the modular product of all the original $t'$ numbers. The resulting
height of the tree is $(\lceil \log_{3/2}(t'/3) \rceil + 1)$ modules.

As the parallel modular additions proceed by layers, all previous layers
must be maintained in a coherent state, since the modular addition leaves
garbage bits behind. Only at the end of modular multiple addition, after
the final answer $u+v$ is obtained, can all the previous layers be
uncomputed in reverse to free up their ancillae.

\begin{figure*}[htb!]
\centerline{
\includegraphics[width=5.5in]{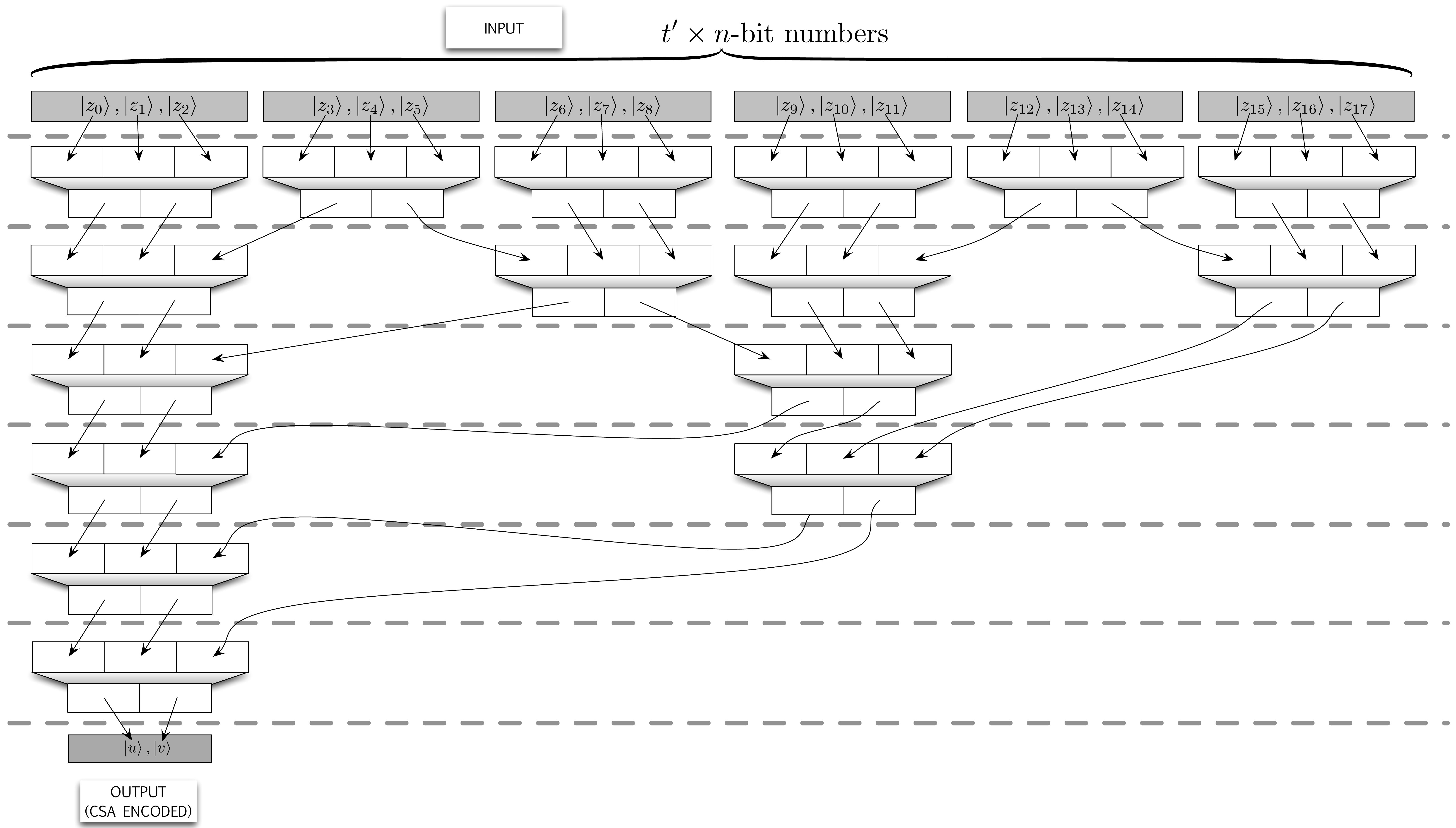}
}
\caption{Modular multiple addition of quantum integers on a CSA tile
architecture for $t'=18$ in a logarithmic-depth tree with height $(\lceil \log_{\frac{3}{2}}(t'/3) \rceil + 1) = 6$. Arrows represent teleportation
in between modules.}
\label{fig:mod-mult}
\end{figure*}

These steps are best illustrated with a concrete
example in Figure \ref{fig:mod-mult}. The module for each CSA tile is
represented by the symbol from Figure \ref{fig:csa-tile-symbol}.
The arrows indicate the
teleportation of output numbers from the source tile to be input numbers
into a destination tile.

Now we can analyze the circuit resources for multiplying $n$-bit
quantum integers, which requires $(t'-2)$ modular additions, for $t'$ from
Equation \ref{eqn:tprime}.
The circuit width is the sum of the $O(n^3)$ ancillae
needed for partial product creation and the ancillae required for $O(n^2)$
modular additions. Each modular addition has width $O(n)$ and depth $O(1)$
from the previous
section. There are
$\lceil \log_{3/2}(n^2 / 3) \rceil +1 $ timesteps of modular addition. Therefore
the entire modular multiplier circuit has depth $O(\log n)$ and width $O(n^3)$.

\subsection{Modular Multiplier Resources}
\label{subsec:mod-mult-resources}.

The circuit depth of the entire modular multiplier is $O(\log n)$:

\begin{equation}
D_{MM} = 1383 \log_2 n + 3930\text{.}
\end{equation}

The module depth is $O(\log n)$:
\begin{equation}
\overline{D}_{MM} = 2\log_2 n + 11\text{.}
\end{equation}

The circuit size is $O(n^3)$:

\begin{eqnarray}
S_{MM} = & (6n + 9)\log_2 n +\\
        & (1152n^3 + 10780n^2 + 17628n + 7082)\text{.}
\end{eqnarray}

The module size is $O(n^3)$:

\begin{equation}
\overline{S}_{MM} = 15n^3 + 127n^2 + 178n + 50{.}
\end{equation}

The circuit width is $O(n^3)$:

\begin{equation}
W_{MM} = 66n^3 + 558n^2 + 870n + 290\text{.}
\end{equation}

The module width is $O(n^2)$:

\begin{equation}
\overline{W}_{MM} = 4n^2 + 28n + 15\text{.}
\end{equation}

\section{Quantum Modular Exponentiation}
\label{sec:modexp}

We now extend our arithmetic to modular exponentiation, which is repeated
modular multiplication controlled on qubits supplied by a phase estimation
procedure.
If we wish to multiply an $n$-qubit quantum input number $\ket{x}$ by
$t$ classical numbers $a^{(j)}$, we can multiply them in series.
This requires depth $O(t\log n)$ in modular multiplication operations.


Now consider the same procedure, but this time each classical number $a^{(j)}$
is controlled on a quantum bit $p_j$. This is a special case of
multiplying by $t$ quantum integers in series, since a classical number
entangled with a quantum integer is also quantum.
It takes the same depth $O(t\log n)$ as the previous case.
%

\begin{figure*}[tb!]
\centerline{
\includegraphics[width=5.5in]{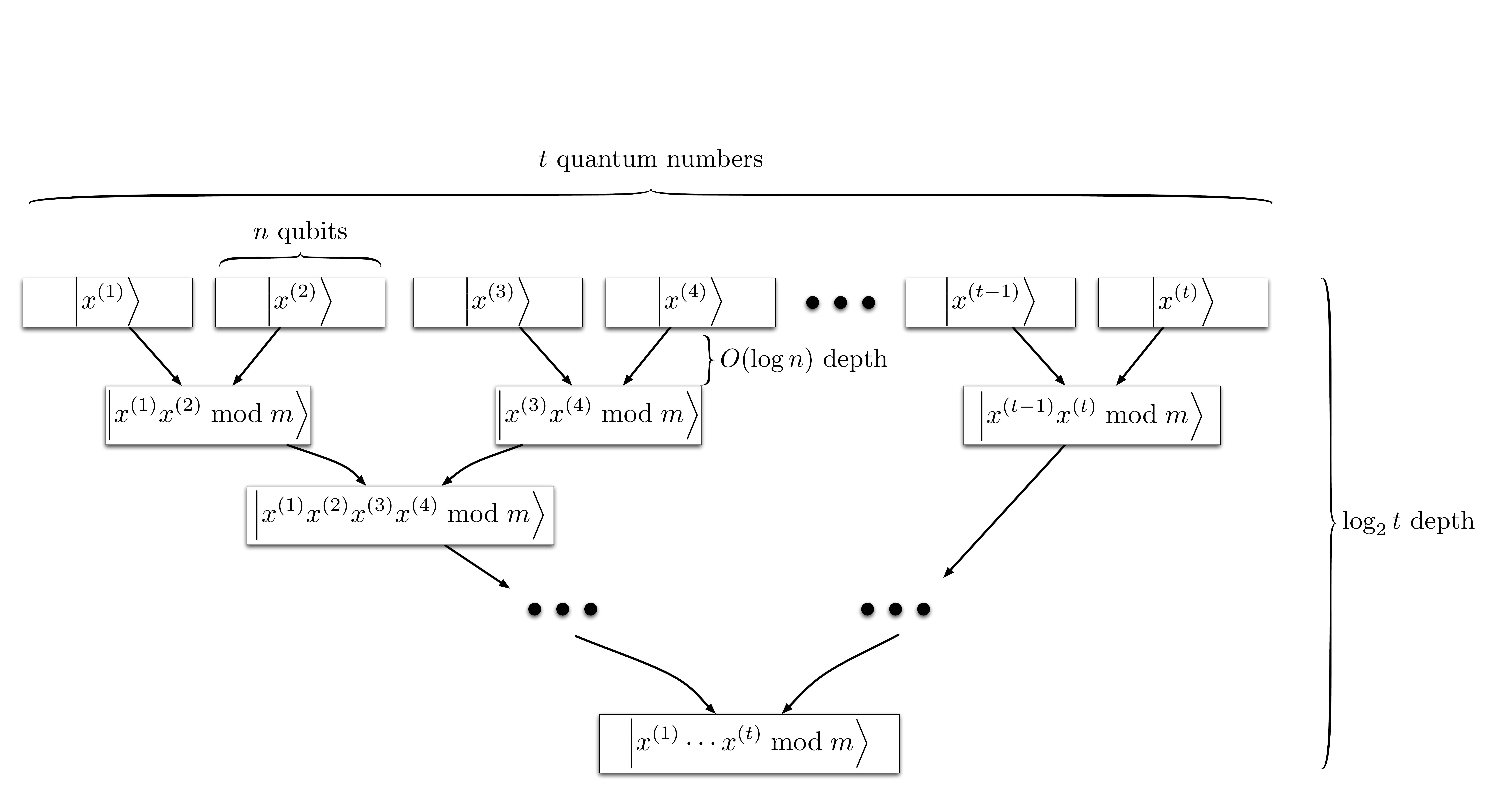}
}
\caption{Parallel modular exponentiation: multiplying $t$ quantum integers
in a $O(\log{(t)}\log{(n)})$-depth binary tree. Arrows indicate modular
multiplication.}
\label{fig:modexp-qq-parallel}
\end{figure*}

Finally, we consider multiplying $t$ quantum integers
$\{x^{(1)}, x^{(2)}, \ldots, x^{(t-1)}, x^{(t)}\}$ in a parallel,
logarithmic-depth binary tree.
This is shown in Figure \ref{fig:modexp-qq-parallel}, where arrows indicate multiplication.
The tree has depth $\log_2(t)$ in modular multiplier operations. Furthermore,
each
modular multiplier has depth $O(\log(n))$ and width $O(n^3)$ for $n$-qubit
numbers. Therefore, the overall depth of this parallel modular exponentiation
structure is $O(\log(t)\log(n))$ with width $O(tn^3)$.
In phase estimation for QPF, it is
sufficient to take $t = O(n)$ \cite{Nielsen2000,Kitaev2002}. Therefore our total depth is
$O(\log^2(n))$ and our total size and total width are $O(n^4)$, as desired. At this point, combined with the parallel phase
estimation procedure of \cite{Kitaev2002}, we have a complete factoring
implementation in our 2D nearest-neighbor architecture in polylogarithmic
depth.

We will now calculate numerical
constants to upper bound circuit resources.

According to the Kitaev-Shen-Vyalyi parallelized phase estimation procedure
\cite{Kitaev2002},
for a constant success probability of $3/4$,
it is sufficient to multiply together $t' = 2867n$ quantum integers,
controlled on the qubits $\ket{p_j}$, in parallel.

In Section \ref{subsec:qcla}, we describe the last step of modular
exponentiation in CSE. In Section \ref{subsec:modexp-resources}, we
state the final circuit resources for the entire modular exponentiation
circuit,
and therefore, our quantum period-finding procedure.

\subsection{Converting Back to a Unique Conventional Number}
\label{subsec:qcla}

The final product of all $t$ quantum integers is in CSE which is not
unique. As stated in Gossett's original paper \cite{Gossett1998}, this
must be converted back to a conventional number using, for example, the
quantum carry-lookahead adder (QCLA) from \cite{Draper2004}. We can convert
this to a nearest-neighbor architecture by using the qubit reordering
construction of \cite{Rosenbaum2012}. We now compute the resources
needed for this last step.

To add two $(n+2)$-bit numbers in a QCLA, we have a circuit width of
$k = (4(n+2) - 2\log_2 n - 1)$. The depth is at most $4\log_2 n +2$ gates,
and some of them act on qubits that are not nearest-neighbors. Therefore,
we add in between each gate a reordering circuit that takes $k^2$
(reusable) ancillae
qubits and uses two rounds of constant-depth teleportation to rearrange
the qubits into a new order where all the gates are nearest-neighbor.
Adding in the teleportation circuit resources from Table \ref{tab:cd-resources},
we can calculate the following resources.

The circuit depth is $O(\log n)$:

\begin{equation}
56\log_2 n + 28\text{.}
\end{equation}

The circuit size is $O(n^2 \log n)$:

\begin{eqnarray}
96 \log_2^3 n & - & (384n + 624)\log_2^2 n \nonumber \\
              & + & (384n^2 + 1152n + 840) \log_2 n \nonumber \\
              & + & (192n^2 + 672n + 588)\text{.}
\end{eqnarray}

The circuit width is $O(n^2)$:

\begin{equation}
4 \log_2^2 n - (16n + 30)\log_2 n + 16n^2 + 60n + 56\text{.}
\end{equation}

\subsection{Circuit Resources for Modular Exponentiator}
\label{subsec:modexp-resources}

This leads to the following circuit resource upper bounds for a modular exponentiator. Therefore, these are the total resources for running a
single round of parallel QPF as part of Shor's factoring algorithm.

The circuit depth is $O(\log^2 n)$:

\begin{equation}
D_{ME} = 1383\log_2^2(n) + 21253\log_2(n) + 49095\text{.}
\end{equation}

The module depth is $O(\log n)$:

\begin{equation}
\overline{D}_{ME} = 3\log_2 n + 24
\end{equation}

The circuit size is $O(n^4)$:

\begin{eqnarray}
S_{ME} & = & 96 \log_2^3 n + \nonumber \\
       & - & (384n + 624)\log_2^2 n \nonumber \\
       & + & (384n^2 + 1152n + 840) \log_2 n \nonumber \\
       &   & 3302324 n^4 + 30900797 n^3 + 50521837 n^2  + 20284306 n + 
  -6494\text{.}
\end{eqnarray}

The module size is $O(n^2)$:

\begin{equation}
\overline{S}_{ME} = 5749n^2 + 8725n +175\text{.}
\end{equation}

The circuit width is $O(n^4)$:

\begin{equation}
W_{ME} = 94598n^4 + 799749 n^3 + 1246692 n^2 + 415222 n - 145\text{.}
\end{equation}

The module width is $O(n)$:

\begin{equation}
\overline{W}_{ME} = 1434n\text{.}
\end{equation}

\section{Asymptotic Results}
\label{sec:results}

The asymptotic resources required for our approach,
as well as the resources for other nearest-neighbor approaches,
are listed in Table \ref{tab:results},
where we assume a fixed constant error
probability for each round of QPF. Not all resources are
provided directly by the referenced source.

Resources in square brackets
are inferred using Equation \ref{eqn:depth-width}.
These upper bounds are correct,
but may not be tight with the upper bounds
calculated by their respective authors.
In particular, a more detailed analysis
could give a better upper bound for circuit size than the
depth-width product. Also note that the
work by Beckman et al. \cite{Beckman1996} is unique in that it uses
efficient multi-qubit gates inherent to linear ion trap technology which at first
seem to
be more powerful than \textsc{1D NTC}. However, use of these gates does not result in an
asymptotic improvement over \textsc{1D NTC}.

We achieve an exponential
improvement in nearest-neighbor circuit depth (from quadratic to polylogarithmic)
with our approach at the cost of a polynomial increase in
circuit size and width. Similar depth improvements at the cost of width increases can be achieved using the modular multipliers
of other factoring implementations
by arranging them in a parallel modular exponentiator.
Our approach is the first implementation for factoring on \textsc{2D NTC},
augmented with a classical controller and parallel, communicating
modules (\textsc{2D CCNTCM}).
\begin{table}[htb!]
\begin{center}
\begin{tabular}{|c|c|c|c|c|}
\hline
Implementation             & Architecture      & Depth   & Size   & Width     \\
\hline
Vedral, et al. \cite{Vedral1996}   & \textsc{AC}      & $[O(n^3)]$ & $O(n^3)$    & $O(n)$ \\
Gossett \cite{Gossett1998}                   & \textsc{AC}       & $O(n \log n)$  & $[O(n^3\log n)]$  & $O(n^2)$  \\
Beauregard \cite{Beauregard2002}                & \textsc{AC}       & $O(n^3)$      & $O(n^3 \log n)$ & $O(n)$ \\
Zalka \cite{Zalka1998}                     & \textsc{AC}       & $O(n^2)$      & $[O(n^3)]$ & $O(n)$     \\
Takahashi \& Kunihiro \cite{Takahashi2006}     & \textsc{AC}       & $O(n^3)$      & $O(n^3\log n)$ & $O(n)$ \\
Cleve \& Watrous \cite{Cleve2000}           & \textsc{AC}       & $O(\log^3 n)$ & $O(n^3)$ & $[O(n^3 / \log^3n)]$ \\
\hline
Beckman et al. \cite{Beckman1996} & \textsc{Ion trap}   & $O(n^3)$ & $O(n^3)$ & $O(n)$\\
\hline
Fowler, et al. \cite{Fowler2004} & \textsc{1D NTC}   & $O(n^3)$ & $O(n^4)$ & $O(n)$\\
Van Meter \& Itoh \cite{VanMeter2006} & \textsc{1D NTC}   & $O(n^2 \log n)$ & $[O(n^4\log n)]$ & $O(n^2)$\\
Kutin \cite{Kutin2006}                     & \textsc{1D NTC}   & $O(n^2)$ & $O(n^3)$ & $O(n)$\\
\hline
Current Work               & \textsc{2D CCNTCM}   & $O(\log^2{n})$ & $O(n^4)$ & $O(n^4)$   \\
\hline
\end{tabular}
\end{center}
\caption{Asymptotic circuit resource usage for quantum factoring of an $n$-bit number.}
\label{tab:results}
\end{table}

\section{Conclusions and Future Work}
\label{sec:conclude}

In this paper, we have presented a 2D architecture for factoring on a quantum
computer using a model of nearest-neighbor, concurrent two-qubit
interactions, a classical controller, and communication between
independent modules. We call this new model
\textsc{2D CCNTCM}.
Using a combination of algorithmic
improvements (carry-save adders and parallelized phase estimation)
and architectural improvements (irregular two-dimensional layouts,
constant-depth communication, and parallel modules), we conclude
that we can run
the central part of Shor's factoring algorithm (quantum period-finding)
with asymptotically smaller depth than previous implementations.


A natural extension of the current work is to improve its
depth to be sub-logarithmic on 2D CCNTCM using the approach outlined in
\cite{Hoyer2002,Siu1993}, generalizing the carry-save adder to
a block-save adder using threshold gates. It would also be
beneficial to determine lower bounds for the
time-space tradeoffs involved in Shor's factoring algorithm.
These results would tell us whether we have found an
optimal nearest-neighbor circuit.

\section*{Acknowledgements}
\noindent
The authors wish to thank Aram Harrow, Austin Fowler, and David Rosenbaum for
useful discussions.
P. Pham conducted the factoring part of this work during
an internship at Microsoft Research.
He also acknowledges funding of the architecture and layout portions
of this work from
the Intelligence Advanced Research Projects Activity
(IARPA) via Department of Interior National Business Center contract
number D11PC20167. The U.S. Government is authorized to reproduce and
distribute reprints for Governmental purposes notwithstanding any
copyright annotation thereon. Disclaimer: The views and conclusions
contained herein are those of the authors and should not be
interpreted as necessarily representing the official policies or
endorsements, either expressed or implied, of IARPA, DoI/NBC, or the
U.S. Government.

\bibliography{PhamSvore_arXiv}
\bibliographystyle{ieeetr}

\end{document}